\documentclass[journal, 12pt, onecolumn, draftcls]{IEEEtran}

\ifCLASSINFOpdf
\else
\fi
\usepackage{xcolor}
\usepackage{algorithmic}
\usepackage{array}
\usepackage{stfloats}
\usepackage{cite}
\usepackage{graphicx}
\graphicspath{{Figures/}{../Figures/}}
\usepackage{gensymb,bm}
\usepackage[T1]{fontenc}
\usepackage[utf8]{inputenc}
\usepackage{pdfpages}
\usepackage{textcomp}
\usepackage[cmex10]{amsmath}
\usepackage{booktabs}
\usepackage{url}
\usepackage{subcaption}
\usepackage{url}
\usepackage{multicol}
\usepackage{multirow}
\begin{document}

\title{THz Band Channel Measurements and Statistical Modeling for Urban D2D Environments}

\author{Naveed A. Abbasi,
        Jorge~Gomez-Ponce, Revanth Kondaveti,
        Shahid M. Shaikbepari,
        Shreyas Rao,
        Shadi Abu-Surra, Gary Xu, Charlie Zhang,
        and~Andreas~F.~Molisch
\thanks{The work of USC was partly supported by the Semiconductor Research Corporation (SRC) under the ComSenTer program, Samsung Research America and the Foreign Fulbright Ecuador SENESCYT Program.}
\thanks{N. A. Abbasi, J. Gomez-Ponce, R. Kondaveti, S. M. Shaikbepari, S. Rao and A. F. Molisch are with the Ming Hsieh Department of Electrical and Computer Engineering, University of Southern California, Los Angeles, CA, USA. J. Gomez-Ponce is also with the ESPOL Polytechnic University, Escuela Superior Politécnica del Litoral, ESPOL, Facultad de Ingenier\'ia en Electricidad y Computaci\'on, Km 30.5 vía Perimetral, P. O. Box 09-01-5863, Guayaquil, Ecuador. S. Abu-Surra, G. Xu and C. Zhang are with Samsung Research America, Richardson, TX, USA. Corresponding author: Naveed A. Abbasi (nabbasi@usc.edu).}
}
\maketitle

\begin{abstract}
THz band is envisioned to be used in 6G systems to meet the ever-increasing demand for data rate. However, before an eventual system design and deployment can proceed, detailed channel sounding measurements are required to understand key channel characteristics. In this paper, we present a first extensive set of channel measurements for urban outdoor environments that are ultra-wideband (1 GHz 3dB bandwidth), and double-directional where both the transmitter and receiver are at the same height. In all, we present measurements at 38 Tx/Rx location pairs, consisting of a total of nearly 50,000 impulse responses, at both line-of-sight (LoS) and non-line-of-sight (NLoS) cases in the 1-100 m range. We provide modeling for path loss, shadowing, delay spread, angular spread and multipath component (MPC) power distribution. We find, among other things, that outdoor communication over tens of meters is feasible in this frequency range even in NLoS scenarios, that omni-directional delay spreads of up to 100 ns, and directional delay spreads of up to 10 ns are observed, while angular spreads are also quite significant, and a surprisingly large number of MPCs are observed for 1 GHz bandwidth and 13$^{\circ}$ beamwidth. These results constitute an important first step towards better understanding the wireless channel in the THz band.
\end{abstract}

\IEEEpeerreviewmaketitle

\section{Introduction}
A number of new and upcoming applications require ultra-high data rates that are beyond the capabilities of mmWave-based 5G communication systems \cite{5764977,Akyildiz201416,7894280}. To fulfill these requirements, higher frequencies are being investigated because large swaths of unused spectrum are available there. The THz band (0.1-10 THz)\footnote{Some authors prefer the term `THz' to identify with frequencies only beyond 300 GHz and `sub-THz' or `low-THz' for frequencies between 100-300 GHz whereas other authors use the term `THz' for both these cases. For the current work, we will use the term `THz' since it is more widely used in the literature for the band of interest (145-146 GHz).}, especially the frequencies between 0.1-0.5 THz, has been widely explored in this regard, e.g., \cite{Kurner2014,akyildiz2014terahertz,petrov2016terahertz,khalid2017capacity,huq2019terahertz,chen2019survey,rappaport2019wireless}. The recent decision of the US spectrum regulator, the Federal Communication Commission (FCC), to provide experimental licenses in this band further fostered research interest, and this band is widely expected to be an important part of 6G wireless systems \cite{tataria20216g}. 

Before the design of a communication system can commence, the characteristics of the channel in which it will operate need to be understood properly. These key channel characteristics can be found by statistical analysis of channel sounding measurements. Therefore, channel sounding is essentially the first step towards the design and deployment of a wireless system \cite{molisch2012wireless}. Since channel characteristics are highly dependent on the scenarios and environment of the envisioned deployment, the respective channel sounding measurements also need to be conducted in the desired environment. 

Most existing channel measurements in the THz bands are limited to short-distance indoor channels \cite{priebe2010measurement,6898846,6574880,khalid2019statistical,abbasi2020channel,xing2021millimeter}, usually because of inherent limitations of the measurement setup, however, recently there has been some progress on longer distances and outdoor scenarios as well. These include the first long-distance (100 m) double-directional channel measurements for the 140 GHz band, which was reported in 2019 \cite{abbasi2019double,abbasi2020double} by our group, as well as our recent works \cite{abbasi2021ultra,abbasi2021double} where we target the device-to-device (D2D) scenario. Recently \cite{xing2021propagation} reported channel measurements and path loss modeling at 140 GHz over longer channel lengths in an urban scenario, with Tx elevated compared to the Rx. Our current paper aims to expand our previous work \cite{abbasi2020double,abbasi2021ultra,abbasi2021double} by providing an extensive measurement campaign with sufficient points to allow meaningful statistical evaluation. To the best of our knowledge, such a detailed channel measurement campaign for Tx and Rx at a similar height has not been conducted before in the THz band.

This paper presents results for our channel measurements in urban outdoor scenarios. These results are based on ultra-wideband double-directional channel measurements between 145-146 GHz, conducted at 38 different transmitter (Tx) - receiver (Rx) location pairs. 21 of these location pairs represent line-of-sight (LoS) scenarios with direct Tx-Rx distances ranging from 1 m to 100 m, while 17 location pairs are non-line-of-sight (NLoS) cases with direct Tx-Rx distances ranging between 2.4 m and 100 m. Based on the nearly 50,000 directional impulse responses we collected from these measurements, we model the path loss, shadowing, delay spread, angular spread and multipath (MPC) power distribution for both LoS and NLoS cases. Our detailed analysis includes results both for the maximum-power-beam direction and the omni-directional characteristics as well as the distance dependence of the parameters, and the relevant confidence intervals for the various model fits. 

The remainder of this paper is organized as follows. In Section II, we describe the channel sounding setup and the measurement locations. Key parameters of interest and their processing is described in Section III. The results of the measurements and modeling are presented in Section IV. We finally conclude the manuscript in Section V.

\section{Measurement equipment and site}
\subsection{Testbed description}
\begin{figure}[t!]
	\centering
	\includegraphics[width=12cm]{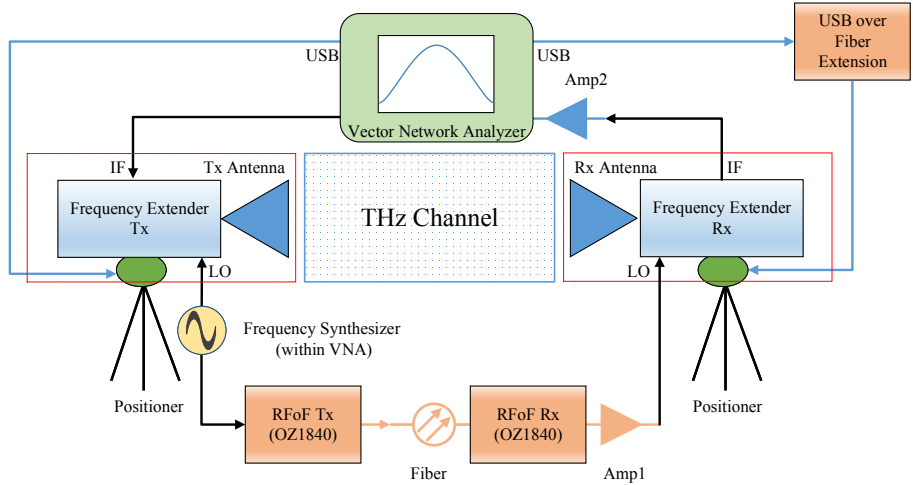}
	\caption{Channel sounding setup.}
	\label{fig:setup}
\end{figure}
\begin{table}[t!]
	\centering
	\caption{Setup parameters.}
	\label{table:parameters}
	\begin{tabular}{|l|l|l|}
		\hline
		\textbf{Parameter}              & \textbf{Symbol}   & \textbf{Value} \\ \hline\hline
		\textit{Frequency points per sweep}     & $N$                 & 1001           \\
		\textit{Tx/Rx height}     & $h_{Tx/Rx}$                 & 1.6 m           \\
		\textit{Start frequency}        & $ f_{start} $     & 145 GHz          \\
		\textit{Stop frequency}         & $ f_{stop} $      & 146 GHz         \\
		\textit{Bandwidth}              & $BW$              & 1 GHz         \\	
		\textit{IF bandwidth}              & $IF_{BW}$              & 10 KHz         \\		
		\textit{THz IF}                   & $ f_{THz IF} $ & 279 MHz          \\
		\textit{Antenna 3 dB beamwidth}   & $HPBW$        & 13$^{\circ}$ \\
		\textit{Tx rotation range}   & $\phi_{Tx}$        & [0$^{\circ}$,360$^{\circ}$]           \\ 
		\textit{Tx rotation resolution}   & $\Delta \phi_{Tx}$        & 10$^{\circ}$           \\
		\textit{Rx Az rotation range}   & $\phi_{Rx}$        & [0$^{\circ}$,360$^{\circ}$]           \\
		\textit{Rx Az rotation resolution}   & $\Delta \phi_{Rx}$        & 10$^{\circ}$           \\ \hline
	\end{tabular}
\end{table}

Our work is based on a custom frequency-domain channel measurement setup shown in Fig. \ref{fig:setup}. The basic principle of this channel sounder revolves around the frequency extension of a vector network analyzer (VNA) signal into the THz domain by means of frequency multipliers. Specifically, we use a Keysight VNA (PNAX N5247A), with VNA extenders from Virginia Diodes (WR-5.1VNAX) where we use the ``high sensitivity" waveguide option to improve the received SNR. As outlined in Table 1, we use an IF bandwidth of $10$ kHz, which provides good SNR while keeping the duration of a measurement sweep less than the time for the mechanical rotation of the horn (see below), so that an increase in that bandwidth would not lead to a significant acceleration of the measurement. The sweep uses 1001 measurement points over a 1 GHz bandwidth, thus enabling to measure up to 1 $\mu$s excess delay without aliasing, corresponding to 300 m maximum excess runlength of the multipath, which is sufficient for the considered environments. 

While VNA-based setups have a long tradition for THz channel measurements, a major challenge has been the fact that Tx and Rx are in a single casing and the maximum separation of the antennas is determined by the admissible length of the cables from the VNA to the extenders (which are placed together with the antennas), which - due to the high cable losses at mmWave frequencies - are usually less than 10 m. In order to overcome this problem, we introduced, in  \cite{abbasi2020double}, a RF-over-fiber (RFoF) link to allows the extension of the distance to the 100 m of interest in our case and beyond. The main difference between the design in the current paper and that in our previous work \cite{abbasi2020double} is that we currently use an integrated RFoF unit to improve the robustness of the design. Due to the measurement principle of mechanically rotating antennas, every measurement lasted for several hours, therefore, it was ensured that the scenarios remain static/quasi-static during this time. 

To determine the double-directional properties of the channel, we measured the transfer functions for different pairs of antenna orientations. Both Tx and Rx are horn antennas with a 3 dB (Full width half maximum) beamwidth of 13 degrees. Due to the equal height of Tx and Rx, it can be anticipated that propagation is happening dominantly in the horizontal plane, therefore, no elevation sweeps were performed. 
The positioners (rotors) were set such that the angle of 0$^\circ$ for both the Tx and the Rx corresponds to the LoS for all points. The Tx scans over a single elevation cut ($\theta_{Tx}=90^\circ$; $\theta_{Tx}=0^\circ$ is defined to be in the zenith) from 0$^\circ$ to 360$^\circ$ with a $10^\circ$ angular resolution. For the Rx, a single elevation  ($\theta_{Rx}=90^\circ$; $\theta_{Rx}=0^\circ$ is again defined to be in the zenith) is taken and a full azimuth scan is done similar to Tx. Since we have a single elevation cut at the Tx and Rx, the use of terms $\theta_{Tx},\theta_{Rx}$ is redundant and they are not used further. Finally, both the Tx and Rx are kept at the same height (1.6 m), corresponding, e.g., to device-to-device (D2D) scenarios. The measurements extended over many days. In the beginning of each measurement day, a calibration of the VNA, as well as an over-the-air (OTA) calibration with the Tx and Rx with LoS in close proximity, was performed. More details of this setup are covered in \cite{abbasi2020double,abbasi2021double,abbasi2021ultra}.

It is pertinent to note that the current setup provides high phase stability that allows to not only conduct Fourier analysis of on our measurements but also enables high resolution parameter extraction (HRPE) algorithms to operate on the measurements. Although beyond the scope of the current paper, we note that HRPE can improve the resolution of the results and therefore provide more accurate modeling. 

\subsection{Measurement locations}

As discussed earlier, selection of measurement locations is important for any measurement campaign and should be representative of the scenario of interest. Our investigation is focused on outdoor urban D2D scenarios, therefore, we conducted measurement campaigns in two relevant environments: (i) an outdoor courtyard surrounded by buildings and (ii) an urban outdoor crossroad. We describe both these locations in more detail in the following.
\subsubsection{Outdoor courtyard ("Epstein Family Plaza/Vivian Hall of Engineering")}
\begin{figure}[ht]
        \centering
        \includegraphics[width=1\columnwidth]{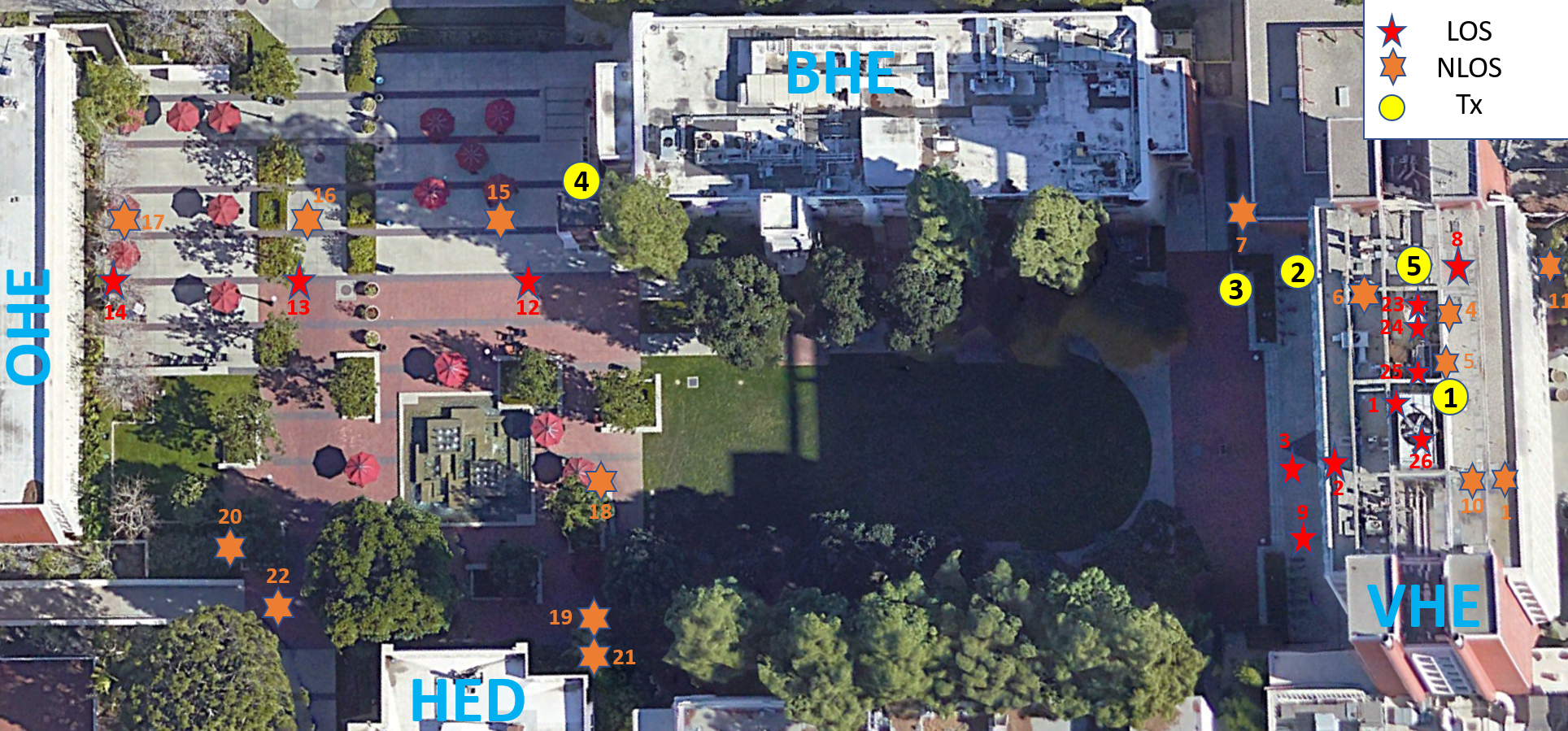}
        \caption{Scenario map for Epstein Family Plaza and VHE.}
        \vspace*{0mm}
        \label{fig:VHE}
\end{figure}

This environment is a quadrangle courtyard (quad) in front of Vivian Hall of Engineering (VHE) on the USC University Park Campus, Los Angeles, CA, USA, see Fig. \ref{fig:VHE}. The Eastern (rightmost on the map) part is an open area with concrete pillars having interspersed benches in the middle and building walls and doors on both sides. To the left is an L-shaped courtyard. The area is partly delimited by multi-story buildings with concrete walls and glass windows. In one part of the L, near the buildings on both sides, are trees with big canopies, as well as lampposts. The other part of the L (left) has fewer trees and includes a large sitting area with tables, chairs, umbrellas, and a water fountain. In this scenario we used 5 different locations for the Tx and 26 locations for the Rx (12 LoS, and 14 NLoS), as  shown in Fig. \ref{fig:VHE}.

Tx1 has 3 LoS points (Rx1-3) and 4 NLoS points (Rx4-7) at the entrance of VHE. Tx2 has 2 LoS locations (Rx8,9) and 2 NLoS points (Rx10,11). Tx3 has 3 LoS points on a linear route crossing the quad (Rx12-14) and 3 NLoS points (Rx15-17) on a linear route blocked by a building corner and foliage. Tx4 has 5 NLoS points (Rx18-22) that are all blocked by foliage, a water fountain and objects in the area. Finally, Tx5 has corresponding receiver locations on a linear route with 4 LoS points (Rx23-26) at VHE building entrance that were previously detailed in \cite{abbasi2021double} \footnote{The measurements in \cite{abbasi2021double} were conducted in the 140-141 GHz band, however, they are included for the current statistical analysis since the relative difference in frequency between these and the rest of the measurements detailed here is very small.}.

\subsubsection{Outdoor crossroad ("Childs- and Watt Way Intersection")}
\begin{figure}[ht]
        \centering
        \includegraphics[width=1\columnwidth]{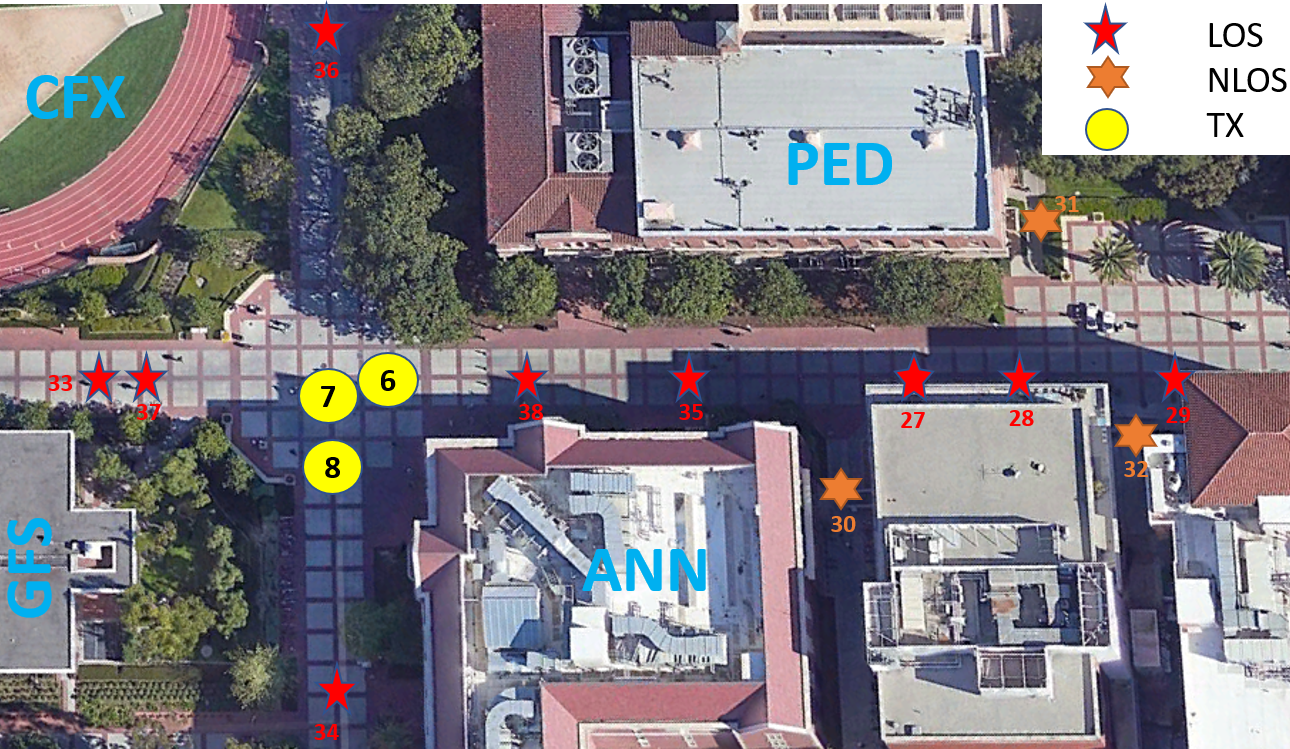}
        \caption{Childs and Watt Way intersection scenario.}
        \vspace*{0mm}
        \label{fig:Anneberg}
\end{figure}

The second scenario is on the intersection between Childs Way and Watt Way also at the USC University Park Campus, see Fig. \ref{fig:Anneberg}. Going north on Watt Way is Cromwell Field (CFX) on the left with sparse trees and a metal fence boundary. Physical Education Building (PED), which has some trees outside it, is opposite to CFX. Looking south on Watt Way, the road lies between Wallis Annenberg Hall (ANN) and Grace Ford Salvatori Hall (GFS). Looking east on Childs Way, we have a street canyon between PED and ANN with some trees on the PED side of the road. After PED, an open area, Associate Park, can be observed with interspersed trees and benches. Finally, on the west side of Childs Way, the road goes between CFX and GFS where the GFS side has larger trees, big glass windows and glass doors whereas the CFX side has smaller sparse trees and a metal fence. For this scenario, 3 Tx locations and 12 Rx locations were selected (9 LoS, 3 NLoS) in total. Fig. \ref{fig:Anneberg} shows the scenarios with the measurement locations.

Tx6 has 3 LoS points (Rx27-29) on linear route and 3 NLoS points blocked by the building corners (Rx30-32). Tx7 was placed in the middle of the crossroad to have 4 LoS points (Rx33-36), one on each of the four sides. Finally, Tx8 has 2 LoS points (Rx37-38) diagonally with respect to its position and the crossroad. Details of all the links for both the sites are provided in Table \ref{tab:dist_Tx-Rx}.
\begin{table}[ht]
\centering
\caption{Description of Tx-Rx links and their respective direct distances.}
\label{tab:dist_Tx-Rx}
\begin{tabular}{|c|c|c|c|c|}
\hline 
            \textbf{Tx identifier}          & \textbf{LoS Rx identifier}         & $\mathbf{d_{LoS}}$ \textbf{(m)}    & \textbf{NLoS Rx identifier}  & $\mathbf{d_{NLoS}}$ \textbf{(m)} \\ \hline \hline
\multicolumn{1}{|l|}{\textbf{$Tx_1$}}           & 1-3               & 2.5, 10, 14             & 4-7       & 2.25, 7, 10, 27        \\ \hline
\multicolumn{1}{|l|}{\textbf{$Tx_2$}}           & 8,9                     & 24, 35                    & 10-11            & 25,23               \\ \hline
\multicolumn{1}{|l|}{\textbf{$Tx_3$}}           & 12-14             & 60, 80, 93              & 15-17       & 60, 80, 93            \\ \hline
\multicolumn{1}{|l|}{\textbf{$Tx_4$}}           &             -         &            -           & 18-22 & 25, 35, 35, 40, 40      \\ \hline
\multicolumn{1}{|l|}{\textbf{$Tx_5$}}           & 23-26          & 1, 2, 5, 15              &   -             &         -            \\ \hline
\multicolumn{1}{|l|}{\textbf{$Tx_6$}}           & 27-29 & 60, 80, 98.4 & 30-32       & 60, 80, 97.59           \\ \hline
\multicolumn{1}{|l|}{\textbf{$Tx_7$}}           & 33-36& 25, 35, 40, 40 & - &  -          \\ \hline
\multicolumn{1}{|l|}{\textbf{$Tx_8$}}           & 37,38                & 28, 28                 &          -      &          -           \\ \hline

\end{tabular}
\end{table}

\section{Parameters and processing}

\subsection{Data processing}
We obtained a collection of the frequency scans from the VNA for various Tx-Rx configurations using the measurement setup described in Section II. Each measurement is represented as a three-dimensional tensor $H_{meas}(f,\phi_{Tx},\phi_{Rx};d)$ where $f$ represents the frequency points over the 1 GHz bandwidth (145-145 GHz), $\phi_{Tx}$ and $\phi_{Rx}$ represent azimuth orientation of the Tx and Rx, and $d$ is the Tx-Rx distance. The dimensions of $H_{meas}$ are $N \times N_{Tx} \times N_{Rx}$ where $N$ is the number of frequency points per sweep $(1001)$, and $N_{Tx}$ and $N_{Rx}$ are the number of azimuth directions at the Tx $(36)$ and Rx $(36)$. In order to eliminate the effects of the system (including antennas) transfer function, we perform an OTA calibration $H_{OTA}(f)$ and obtain the calibrated directional channel transfer function $H(f,\phi_{Tx},\phi_{Rx};d)$ by dividing the raw data with the calibration data as $H(f,\phi_{Tx},\phi_{Rx};d)=H_{meas}(f,\phi_{Tx},\phi_{Rx};d)/H_{OTA}(f)$. From this, the directional power delay profile (PDP) is computed as
    \begin{equation}
     P_{calc}(\tau,\phi_{Tx},\phi_{Rx},d)=|\mathcal{F}_{f}^{-1}\{H(f,\phi_{Tx},\phi_{Rx},d)\}|^2,
    \end{equation}
where $\mathcal{F}_{f}^{-1}$ is the inverse fast Fourier transform (IFFT) with respect to $f$.  
Finally, we apply noise and delay gating, similar to \cite{gomez-ponce2020,abbasi2021double}, which is 

\begin{equation}
      P(\tau)=[P_{calc}(\tau): (\tau\leq\tau_{gate})  \land (P_{calc}(\tau)\geq P_{\lambda})]
\end{equation}  
or $0$ if it does not fulfill these conditions. Here $\tau_{gate}$ is the delay gating value selected to avoid using long delay points and points with "wrap-around" effect of the IFFT, and $P_{\lambda}$ is the noise threshold to not count delay bins with noise which could particularly distort delay spread and angular spread. For the current measurements, $\tau_{gate}$ is set to 833.33 ns (corresponding to 250 m excess runlength) and $P_{\lambda}$ is selected to be 6 dB above the noise floor (average noise power) of the PDP.

Additionally, we analyze the channel behavior from an "omni-directional" perspective by reconstructing the omni-directional pattern from a full double-directional capture by an approach similar to similar to \cite{umit,abbasi2021ultra}, i.e., selecting the direction of the highest contribution per delay bin:
\begin{equation}
    P_{\rm omni}(\tau,d)=\max_{\phi_{Tx},\phi_{Rx}}P(\tau,\phi_{Tx},\phi_{Rx},d).
\end{equation}

\subsection{Parameter computation}
\label{sec:par}
Using the directional and omni-directional PDPs described in the previous section, we proceed to compute several condensed parameters that are commonly used to characterize propagation channels, such as path gain, shadowing, delay spread and angular spread.
\subsubsection{Path loss and shadowing}
We compute path loss as the sum of the power on each delay bin in the PDP as described in \cite{molisch2012wireless}.
\begin{equation}
        PL_i(d)=\sum_\tau P_i(\tau,d),
   \end{equation}
where $i$ can denote omni-directional (omni) or strongest beam (best-dir) which is selected as the beam-pair directions with the highest power:

\begin{equation}
    P_{\rm max}(\tau)=P(\tau,\phi_{\hat{i}},\phi_{\hat{j}}); (\hat{i},\hat{j}) = \max_{i,j} \sum_\tau P(\tau,\phi_i,\phi_j).
\end{equation}
Finally, we model path loss on the dB scale using the classical "power law" also know as $\alpha - \beta$ model as 
\begin{equation}
    PL(d)=\alpha+10\beta \log_{10}(d)+\epsilon,
\end{equation}
where $\alpha$ and $\beta$ are the parameters of the linear model, and $\epsilon$ is the random variation of the data with respect to its mean that is commonly modeled as a zero-mean normal distribution $\epsilon \sim N(0,\sigma)$, where $\sigma$ represents the standard deviation of the distribution. These parameters can be obtained by techniques such as maximum likelihood estimation (MLE) or ordinary least squares (OLS) \cite{molisch2012wireless,kartunen_PL}. As has become common in channel models, we separately extract the parameters $\alpha$, $\beta$, $\sigma$ for the ensemble of LoS and of NLoS measurement points.  

One of the challenges in evaluating a measurement campaign is that the density of distances at which measurements are done is not uniform on a logarithmic distance scale (it is generally not uniform on a linear distance scale either).   
For this reason, a weighted regression approach to model path loss data was developed in \cite{kartunen_PL} where  the measurement points are weighted ($w_i$) according to the density distribution along the distance in $log_{10}$ scale. In other words, this method gives more weight to points in area where the density is low. Using this strategy, we compensate the uneven distribution of points when performing the path loss modeling and give equal importance to each distance range. Even though multiple types of weights are shown in \cite{kartunen_PL}, we focus on the case of "equal weights to N bins over $log_{10}(d)$ ($w_i \propto log_{10}(d)$)", because it corresponds to a least square fitting of "dB vs $log_{10}(d)$" as described in \cite{kartunen_PL}. 
\subsubsection{Delay spread}
Delay spread is another key channel parameter that is calculated as the second central moment of the PDP \cite{molisch2012wireless} as
\begin{equation}
    \sigma_\tau=\sqrt{\frac{\sum_\tau P_i(\tau)\tau^2}{\sum_\tau P_i(\tau)} - \left(\frac{\sum_\tau P_i(\tau)\tau}{\sum_\tau P_i(\tau)}\right)^2},
\end{equation}
where $i$ can be "omni" or "max-dir". Note that long-delayed samples with small power can have a disproportionate impact on the delay spread; for this reason, noise thresholding is especially important in the assessment of delay spread. 
\subsubsection{Angular spread}
\label{sect:AS}
To analyze the channel's behavior for systems that have either directional antennas or antenna arrays, we need a quantitative description of the angular dispersion. A starting point is the double-directional angular power spectrum (DDAPS) which shows the distribution of the power as a function of the angle at Tx and Rx (since our measurement setup scans only the horizontal plane, a distinction between angular and azimuthal power spectrum is moot). The DDAPS is computed as 
\begin{equation}
        DDAPS(\phi_{Tx},\phi_{Rx},d)=\sum_\tau P(\tau,\phi_{Tx},\phi_{Rx},d).
\end{equation}
The granularity of the function is directly proportional to angular captures done during the measurement. We note that it is important to perform noise thresholding and delay gating before computing the DDAPS in order to suppress accumulation of noise in the directions in which no significant MPCs occur. 

From the DDAPS, we can obtain the (single-directional) angular power spectrum (APS) at Tx by integrating over $\phi_{Rx}$, and similarly for the APS at the Rx. From these quantities, we proceed to compute the angular spread by applying Fleury's definition \cite{fleury2000first} given as
\begin{equation}
    \sigma^\circ=\sqrt{\frac{\sum_\phi \left|e^{j\phi}-\mu_\phi \right|^2 APS_k(\phi)}{\sum_\phi APS_k(\phi)}},
\end{equation}
where $k$ can be Tx or Rx indicating departure or arrival APS and $\mu_\phi$ can be computed as
\begin{equation}
    \mu_\phi=\frac{\sum_\phi e^{j\phi} APS_k(\phi)}{\sum_\phi APS_k(\phi)}.
\end{equation}
This analysis is directly impacted by the finite horn antenna beamwidth, which will results in larger angular spreads than the channel inherently provides. For instance, a pure LoS case results in a non-zero angular spread due to the finite beamwidth of the horn antenna. Consequently, the results obtained from this analysis are indeed an upper bound for the real angular spreads of the channel.

\subsubsection {Power distribution over MPC}
A very important parameter for the channel analysis is the ratio of the strongest MPC versus the other MPCs present in the channel, as it informs the fading depth. 
We thus define a parameter $\kappa_1$ as 
\begin{equation}
    \kappa_1=\frac{P_i(\tilde{\tau}_1)}{\sum_{\tilde{\tau}=\tilde{\tau}_2}^{\tilde{\tau}_N} P_i(\tilde{\tau})},
\end{equation}
where $i$ can be "omni" or "max-dir", and $\tilde{\tau_k}$ is the location of the $k$-th local maximum of the PDP $P_i(\tilde{\tau})$, ordered by magnitude, so that $\tilde{\tau_1}$ signifies the location of the largest local maximum.   
Please note that $\kappa_1$ is different from "Rice Factor" because in the Fourier analysis it is not possible to differentiate between closely spaced MPCs, so that the local maximum of the PDP is not strictly identical to an MPC. Consideration of Rice factor based on HRPE will be presented in future work.\\
We In the subsequent section, apart from other results we also perform regression analysis for the parameters $\sigma^\circ,\sigma_\tau,\kappa_1$ to observe their behavior with respect to the distance between Tx and Rx. For this case, a linear regression model expressed as $Z=\alpha+\beta \log_{10}(d)$  is used.
\section{Measurement results}
We now turn to the results obtained from the measurements in different scenarios. 
\subsection{Power delay profiles}
\begin{figure}
\centering
    \centering
    \hspace{7mm}
    \includegraphics[width=0.5\columnwidth]{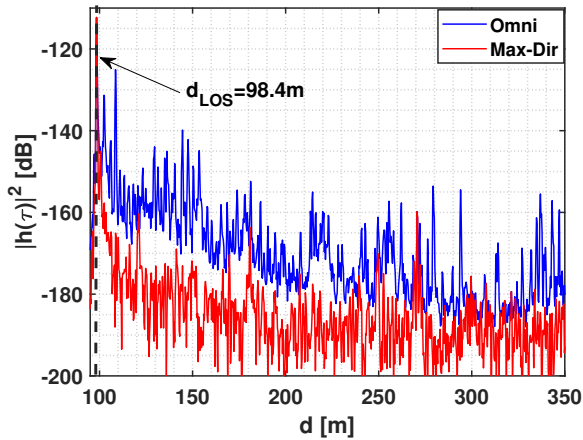}
    \caption{LoS case with $d_{Tx-Rx}=98.4 m$ (Tx6-Rx29).}
    \label{fig:PDP-LOS}
\end{figure}

We first show three sample PDPs, one in a LoS scenario and two for NLoS scenarios. All these cases depict both the "omni" and "max-dir" PDPs as shown in Fig. \ref{fig:PDP-LOS} and Fig. \ref{fig:APS-NLOS}. For the LoS PDP in Fig. \ref{fig:PDP-LOS}, apart from a strong peak at the LoS distance (98.4 m), we observe multiple MPCs due to the interaction of the signal with environmental objects. Clearly, the multi-path structure is much more pronounced in the omni-directional case than in the directional case; this follows intuition, since reflected paths can depart and arrive in a wide variety of angles. The furthest MPCs ($d \geq 200$ m) have power that is more than $40$ dB lower than the strongest component, but closer components may be lower  by only $20$ dB. For the directional case, the extra components are diminished/filtered as a result of the use of horn antennas at both link ends. Additionally, in some cases we observe MPCs before the LoS arrives which are a product of components that arrive later than the maximum measurable distance of the system (1$\mu$s or 300 m) and become visible before the LoS component as a result of the wrap-around effect; this wrap-around is corrected in the figures. 
\begin{figure*}[t!]
\centering
    \begin{subfigure}{0.45\textwidth}
            \centering
            \hspace{7mm}
            \includegraphics[width=1\columnwidth]{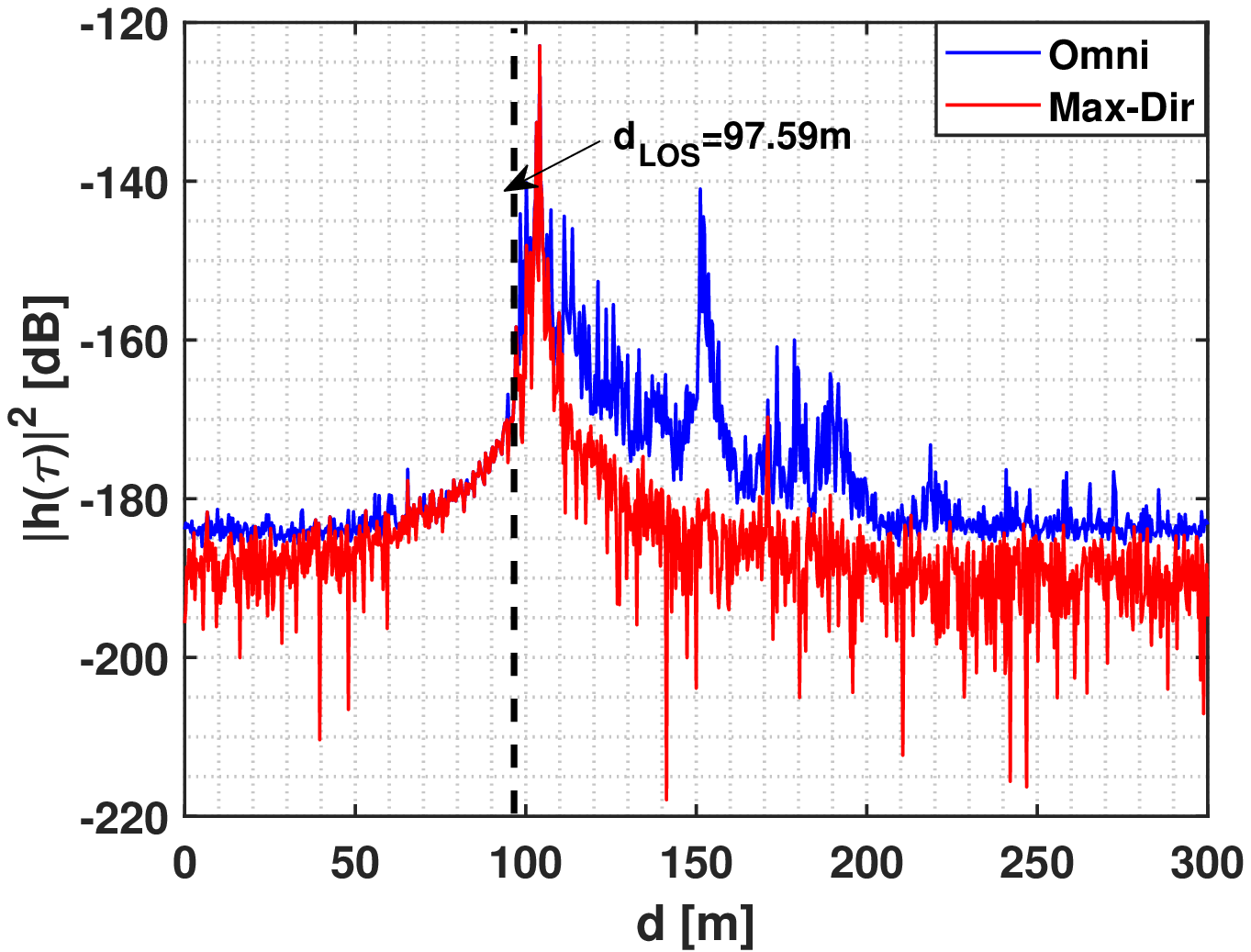}
            \caption{NLoS case with $d_{Tx-Rx}=97.59 m$ (Tx6-Rx32).}
            \label{fig:PDP-NLOS1}
    \end{subfigure}
    \begin{subfigure}{0.45\textwidth}
    \centering
        \centering
        \hspace{7mm}
        \includegraphics[width=1\columnwidth]{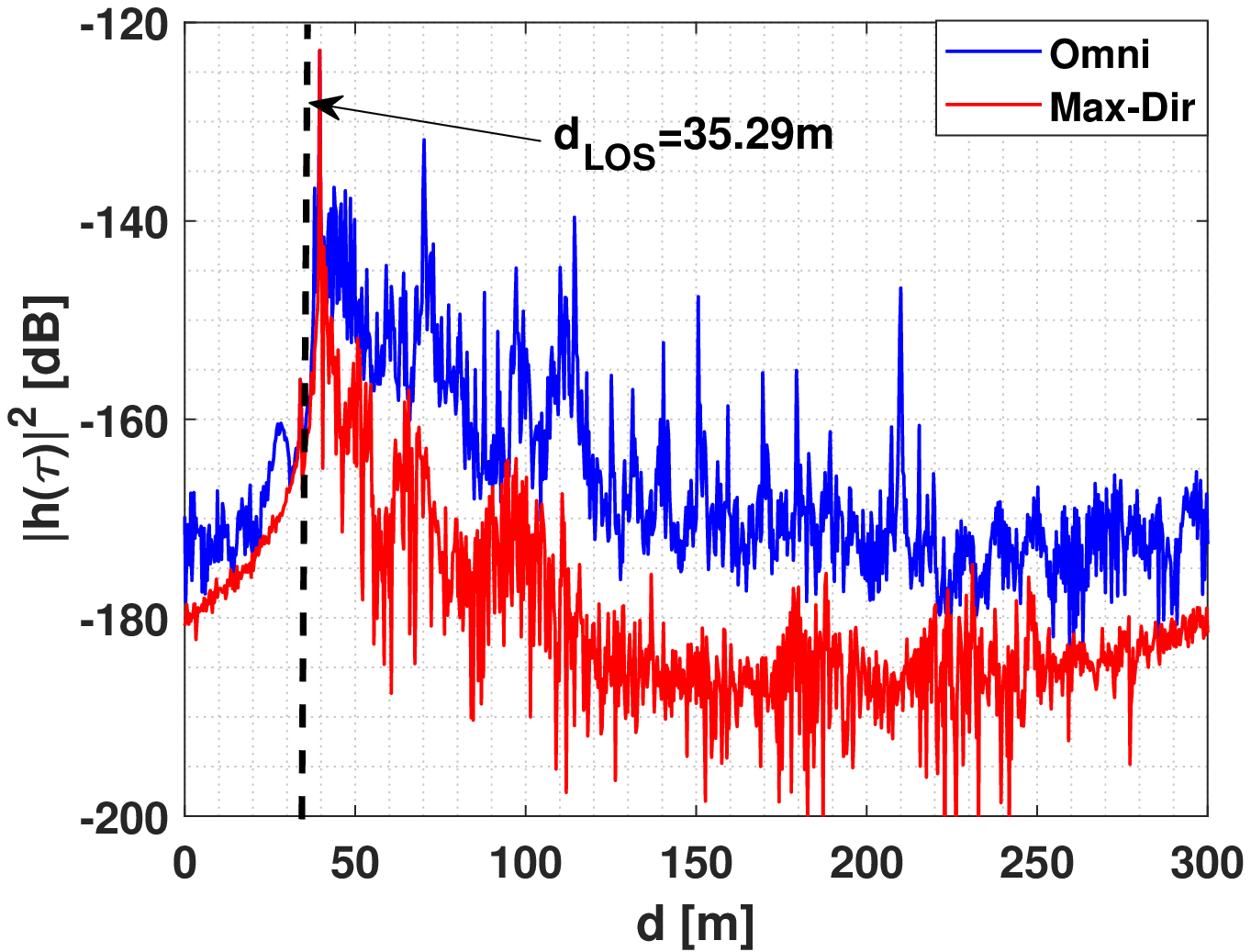}
        \caption{NLoS case with $d_{Tx-Rx}=35.29 m$ (Tx4-Rx20).}
        \vspace*{0mm}
        \label{fig:PDP-NLOS2}
    \end{subfigure}
    \caption{PDP for two sample NLoS measurement cases.}
    \label{fig:PDP-NLOS}
\end{figure*}

The NLoS scenarios are generally richer in terms of MPCs when compared to the LoS case, since the dominant LoS components are missing and therefore, the difference in power between various MPCs is less. Similar to the LoS case, the number of significant MPCs  is larger in the omni-directional case. A high number of MPCs is observed for both the omni-directional and directional cases in Fig. \ref{fig:PDP-NLOS} (b). However, the absence of a dominant component need not be the case for all the NLoS measurements. NLoS scenarios in street canyons, especially observed in the "Childs- and Watt Way Intersection" locations, are noticeably different. One such case is shown in Fig. \ref{fig:PDP-NLOS} (a) where we see a PDP not too different from a LoS case with some very strong components. The primary reason for this is the strong reflections that result from buildings in the scenarios.    
Finally, we note that multipath richness increases with the distance between Tx and Rx since the number of possible paths that a wave can take to get from Tx to Rx increases. This effect was also observed in \cite{abbasi2020double}.
\subsection{Angular power spectrum}
\begin{figure}
\centering
    \centering
    \includegraphics[width=0.5\columnwidth]{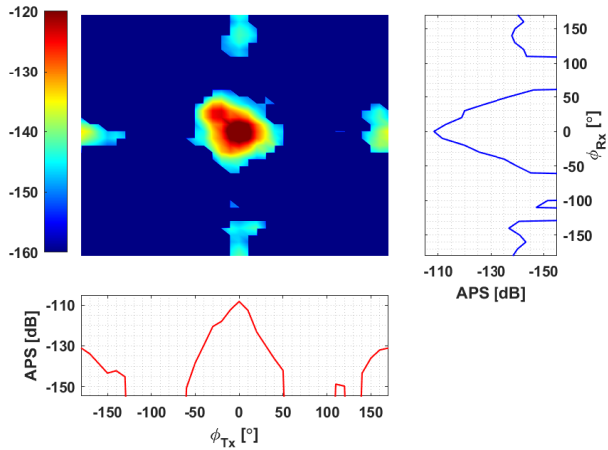}
    \caption{LoS APS for $d_{Tx-Rx}=98.4 m$ (Tx6-Rx29).}
    \vspace*{0mm}
    \label{fig:APS-LOS}
\end{figure}
In terms of APSs for LoS scenarios, we observe a high concentration of power around the LoS direction (i.e., Tx and Rx facing each other, $\phi_{Tx}=\phi_{Rx}=0$) with additional contributions from various directions corresponding to reflections. The concentration of power around the LoS decreases as the distance between Tx and Rx increases.

This is demonstrated by the sample APS case shown in Fig. \ref{fig:APS-LOS}, which exhibits a large concentration of power in the LoS direction and some smaller contributions from the back of both the Tx and the Rx. These contributions are around 20 dB below the LoS components and correspond to reflections from the back - they are not antenna back lobes, which are nearly 30 dB below the main lobe for our antenna.
\begin{figure*}
\centering
    \begin{subfigure}{0.45\textwidth}
         \centering
        \includegraphics[width=1\columnwidth]{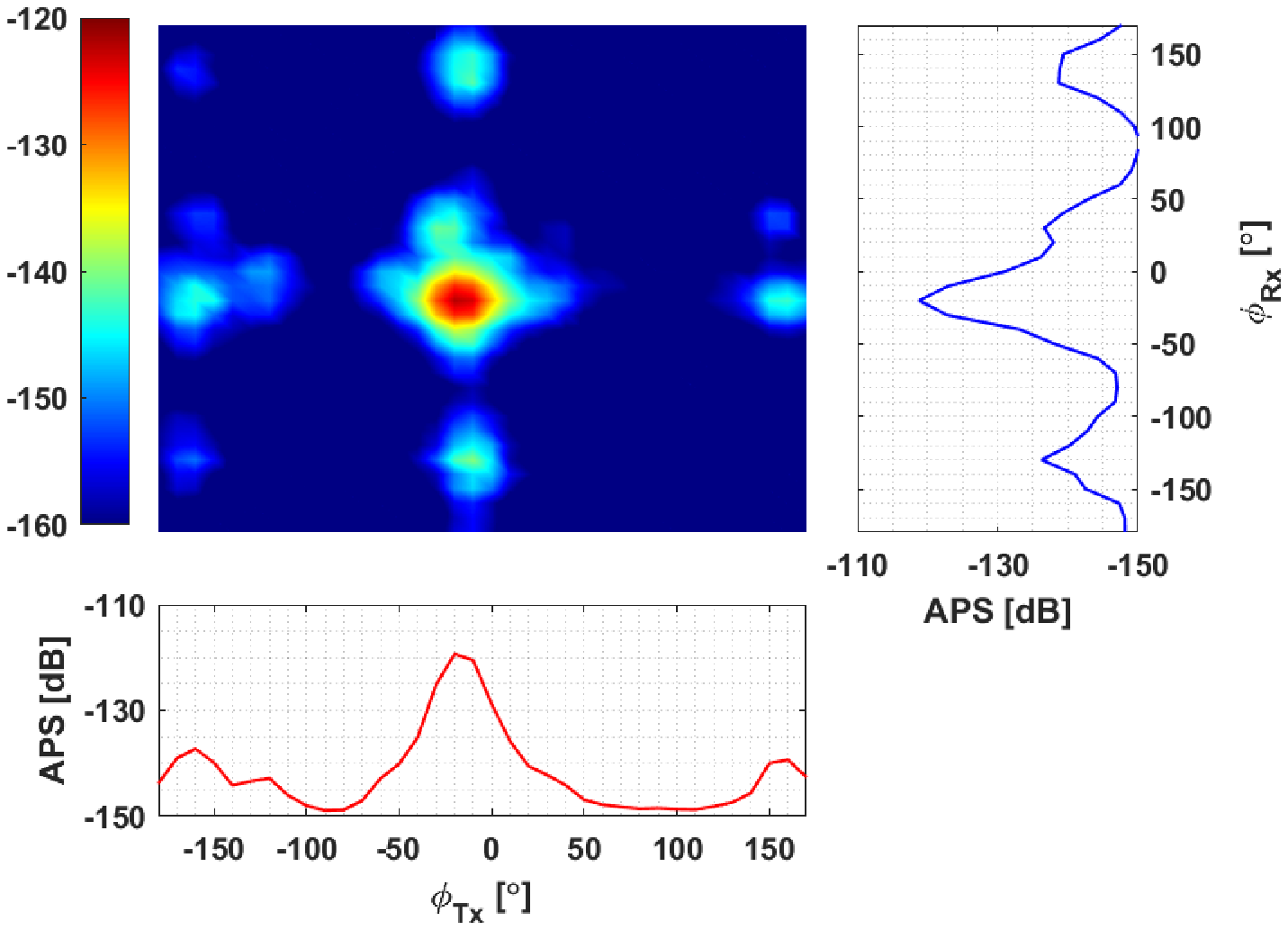}
        \caption{NLoS APS for $d_{Tx-Rx}=97.59$ (Tx6-Rx32).}
        \vspace*{0mm}
        \label{fig:APS-NLOS1}
    \end{subfigure}
    \begin{subfigure}{0.45\textwidth}
         \centering
        \includegraphics[width=1\columnwidth]{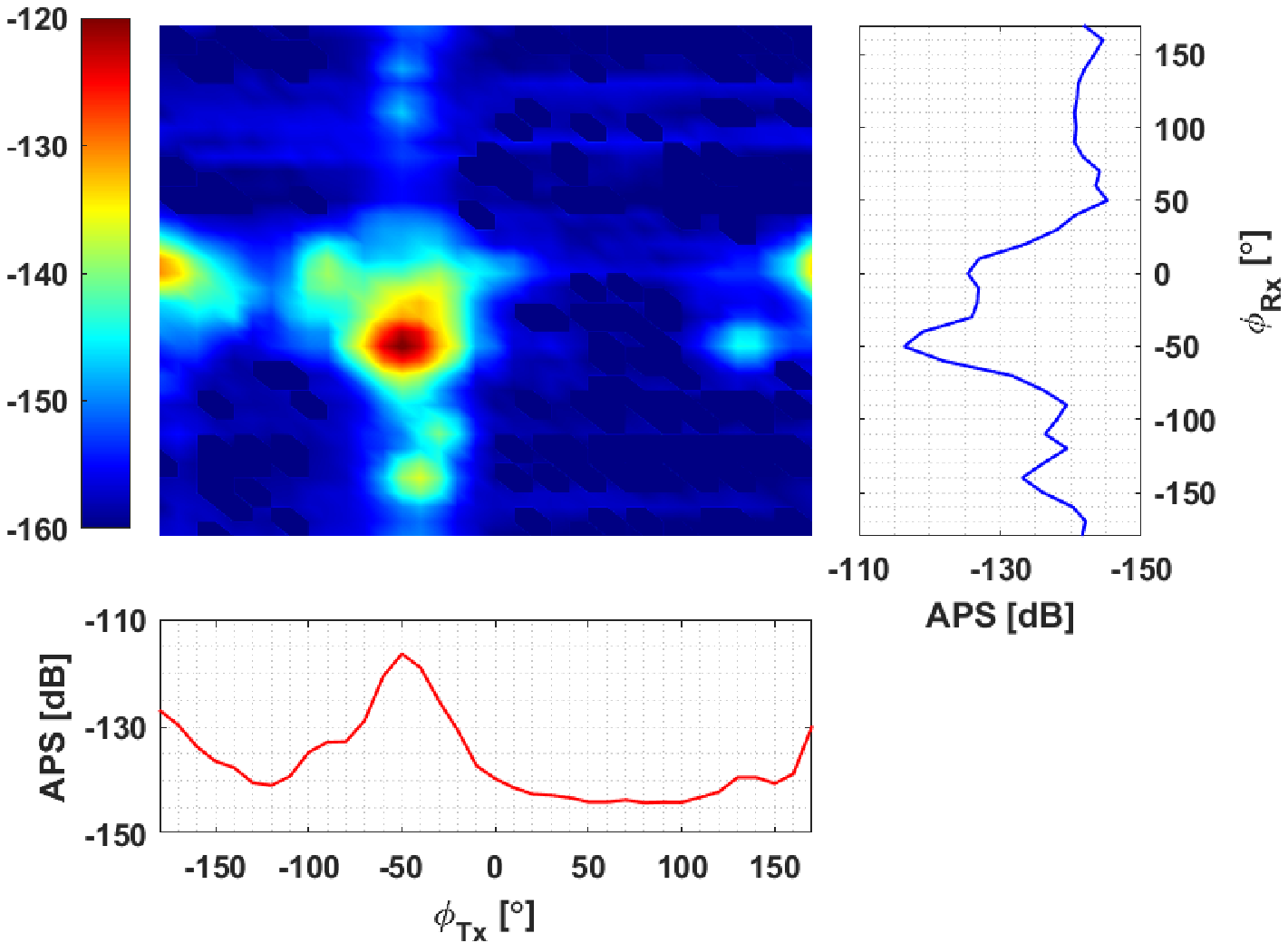}
        \caption{NLoS APS for $d_{Tx-Rx}=35.29 m$ (Tx4-Rx20).}
        \vspace*{0mm}
        \label{fig:APS-NLOS2}
    \end{subfigure}
    \caption{Sample NLoS APSes for two cases.}
    \label{fig:APS-NLOS}
\end{figure*}

For the NLoS case, street canyon scenarios showed peaked (concentrated) APSs, consistent with the concentration of MPCs in the PDPs discussed in the previous subsection. Most other locations showed a considerably larger angular dispersion, and more than one MPC cluster. Representative results for both these cases are shown in Fig. \ref{fig:APS-NLOS}. Fig. \ref{fig:APS-NLOS} (a), shows the street canyon case one major cluster much like a LoS scenario is centered at $\phi_{Tx}=-20^{\circ},\phi_{Rx}=-20^{\circ}$. On the other hand, in Fig. \ref{fig:APS-NLOS} (b), the largest power comes from $\phi_{Tx}=-50^{\circ},\phi_{Rx}=-70^{\circ}$ and additional MPCs clusters with comparable powers can also be observed in other directions.

\subsection{Path loss and shadowing}
\begin{figure*}[t!]
    \begin{subfigure}[b]{0.45\textwidth}
    \centering
        \includegraphics[width=1\columnwidth]{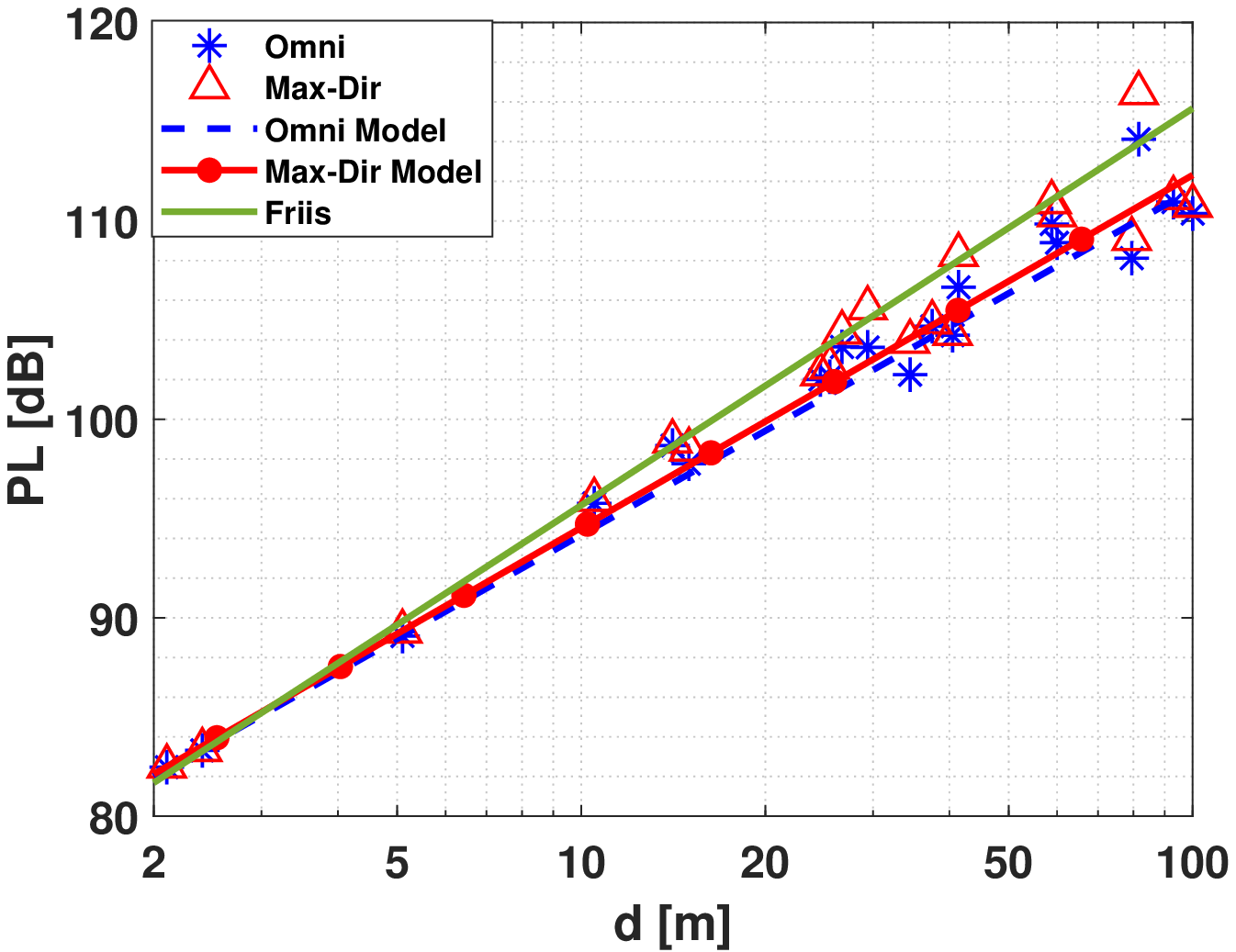}
        \caption{Path loss modeling with $log_{10}(d)$ weighting.}
        \vspace*{0mm}
        \label{PLOSS-LOS}
        \end{subfigure}
    \centering
    \begin{subfigure}[b]{0.45\textwidth}
        \centering
        \includegraphics[width=1\columnwidth]{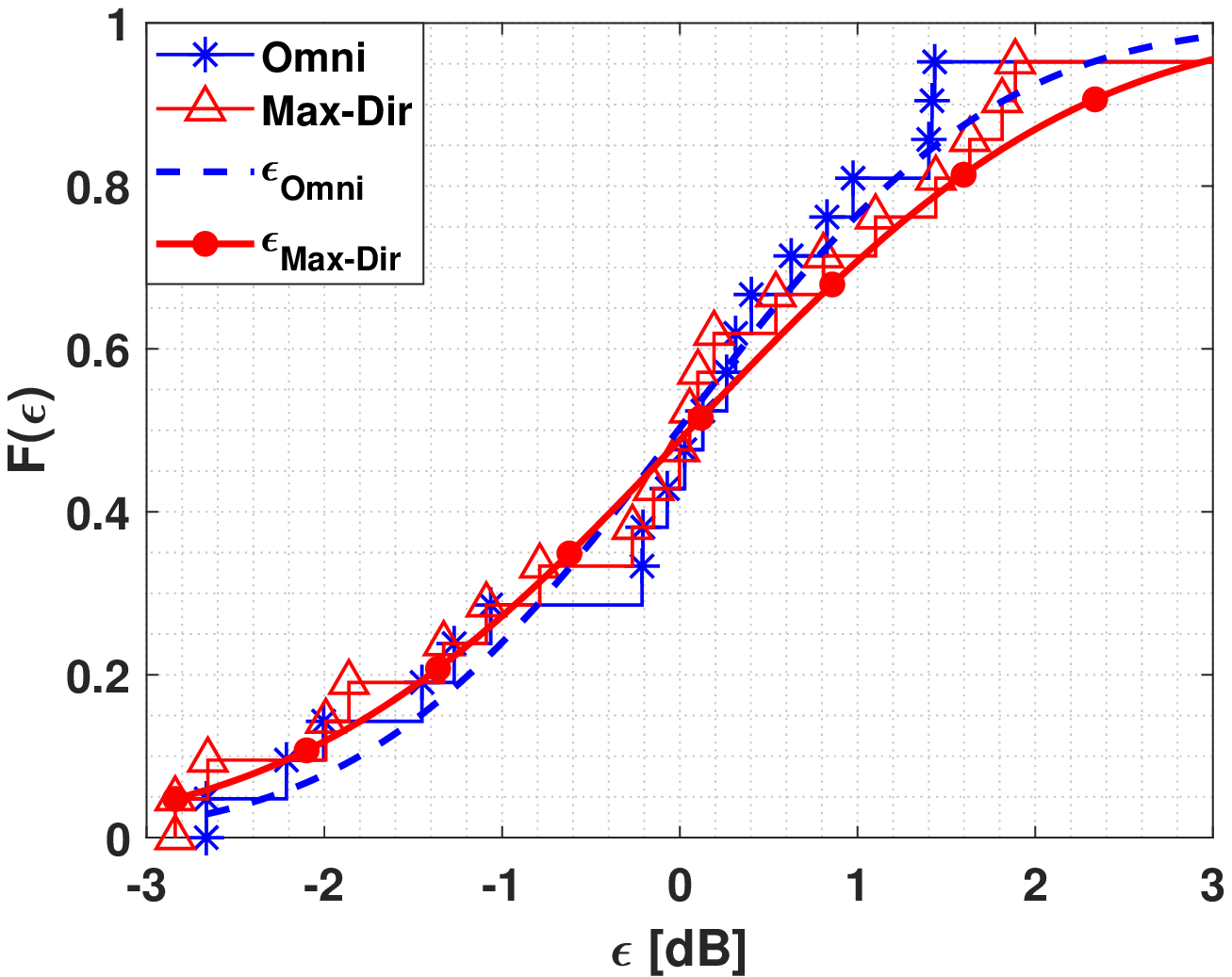}
        \caption{Shadowing.}
        \vspace*{0mm}
        \label{SHA-LOS}
    \end{subfigure}
\caption{Path loss and shadowing models for LoS points.}%
\label{fig:los_PL_SHA}%
\vspace{-0 mm}
\vspace{-5mm}
\end{figure*}

\begin{figure*}[t!]
\centering
\begin{subfigure}[b]{0.45\textwidth}
        \centering
        \includegraphics[width=1\columnwidth]{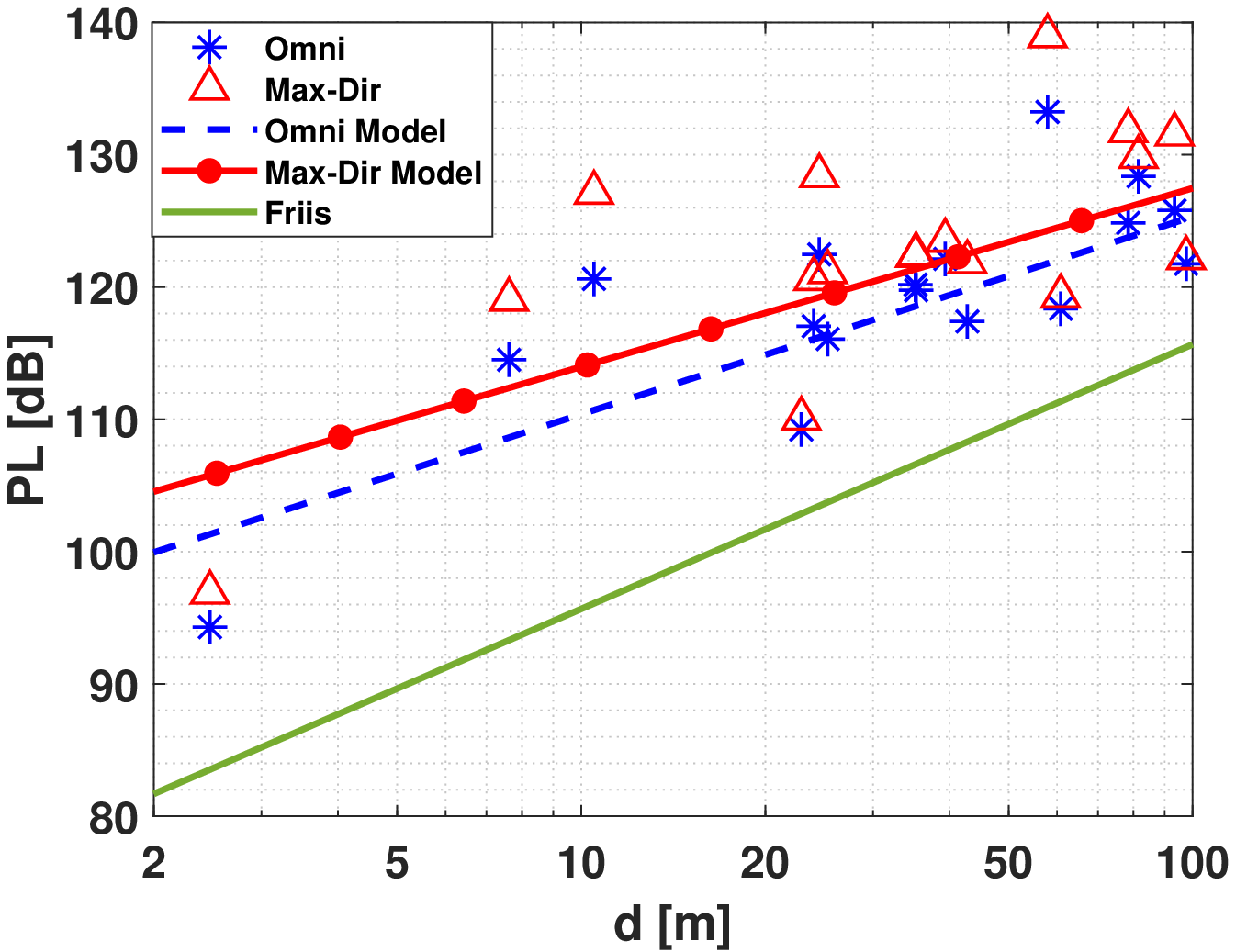}
        \caption{Linear fitting with $log_{10}(d)$ weighting.}
        \label{PLOSS-NLOS}
\end{subfigure}
    \begin{subfigure}[b]{0.45\textwidth}
        \centering
        \includegraphics[width=1\columnwidth]{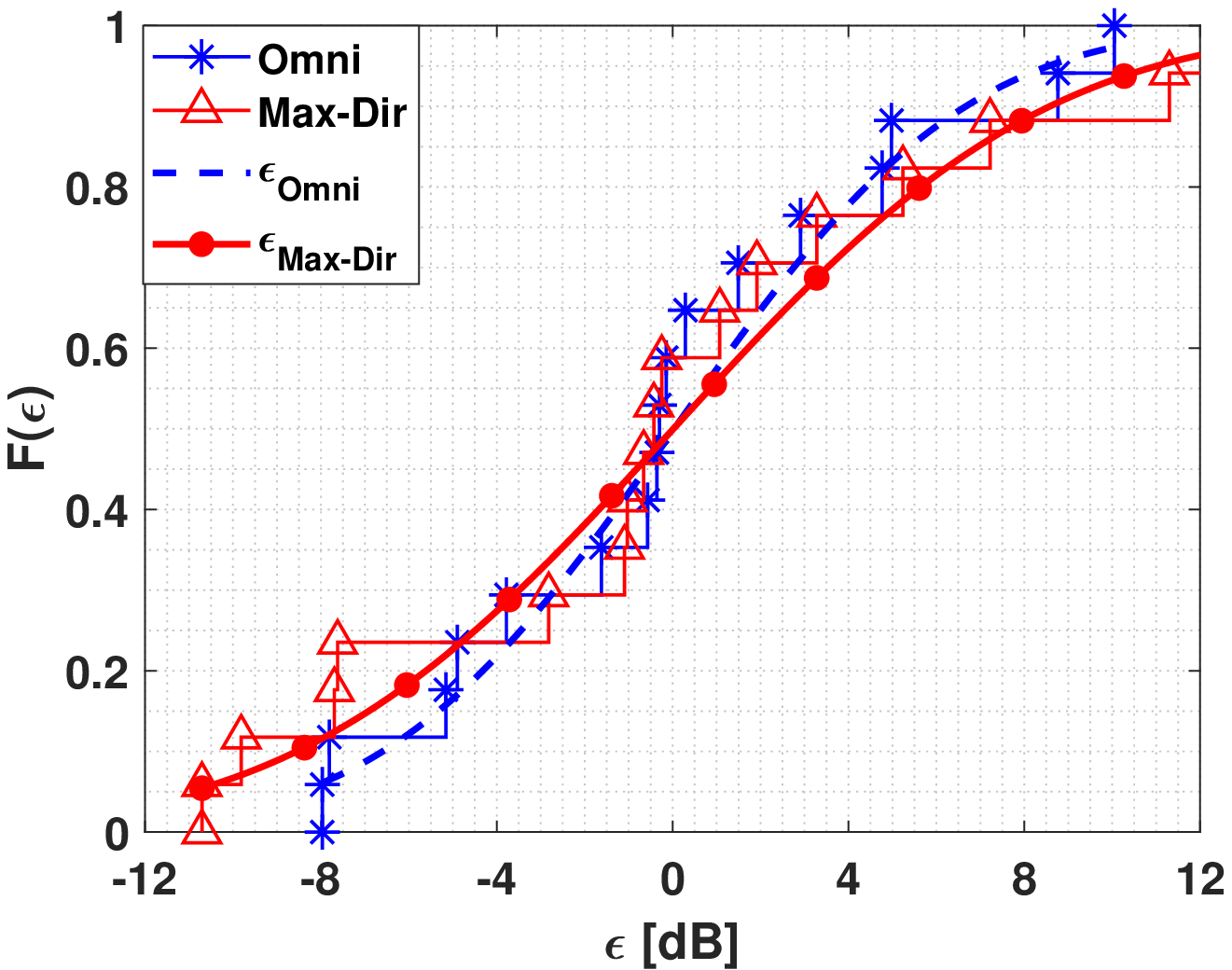}
        \caption{Shadowing.}
        \label{SHA-NLOS}
    \end{subfigure}
\caption{Path loss and shadowing models for NLoS points.}%
\vspace{-0 mm}
\label{fig:nlos_PL_SHA}%
\vspace{-5mm}
\end{figure*}

We proceed with the analysis of the ensemble of locations, starting with path loss. The path loss plots shown in Fig. \ref{fig:los_PL_SHA} (a) and Fig. \ref{fig:nlos_PL_SHA} (a) were done using the weighting regression explained in Section \ref{sec:par}. However, the shadowing results shown are obtained from a regression using the OLS approach, to enforce a "zero-mean" distribution. The difference in mean between the weighted regression and OLS is fairly small; both the weighted fitting and OLS results, for both path loss and shadowing, are detailed in Tables III and VI.  The observations for the LoS measurements shows a close agreement with free space propagation model (FSPL) with small variations around the mean. This can be observed in Fig. \ref{fig:los_PL_SHA} (a) where the omni-directional and directional path loss data exhibit similar trends. The path loss coefficients are between 1.74 and 1.86 for both the directional and the omni-directional case. This is smaller than the FSPL coefficient of 2, which is physically reasonable, as the total received signal consists of several reflected MPCs in addition to the LoS component that provides power decaying with $d^2$. 

The shadowing results shown in Fig. \ref{fig:los_PL_SHA} (b), demonstrate a fairly small shadowing effect; the shadowing standard deviation is $1.73$ dB. Additionally, it can be observed that the variations from the omni-directional case are smaller compared to the directional points which is a result of the larger number of MPCs in the omni-directional PDPs. Detailed results are provided in Tables \ref{tab:PL} and \ref{tab:sha}. We can also see from inspection that the lognormal model that is commonly used for the deviations from the linear fit model provides an excellent approximation to the measured distribution.

The NLoS case exhibits additional attenuation compared to LoS points and FSPL, due to the the impact of blockage, and the losses involved in the additional propagation mechanisms such as reflection and dispersion in the various scenarios. Fig. \ref{fig:nlos_PL_SHA} (a) shows an offset between the FSPL and the measured values between 10 to 15 dB in the directional and omni-directional cases, respectively. The slopes (of path loss as a function of distance) of the directional and omni-directional cases are similar, with the directional path loss slightly larger (since it excludes some significant MPCs). It is noteworthy that the slopes in both cases are around $1.5$, i.e., smaller than in the LoS case, although of course the absolute path loss values are larger than for LoS. We stress that - as in any model based on measurements - the extracted parameters and models are valid only in the range in which measurements were done, i.e., 1-100 m. Disregarding this important caveat would lead to unphysical results at distances larger than $100$ m. 

Shadowing variations are larger in the NLoS case when compared to LoS, and the directional case shows even larger variation compared to the omni-directional case as shown in Fig. \ref{fig:nlos_PL_SHA} (b). This is because of the changes in the strongest MPCs, and the corresponding beam directions that provide the maximum power. More specifically, the shadowing standard deviations $\sigma_{\epsilon_{Omni}}=5.18$ and $\sigma_{\epsilon_{Max-Dir}}=6.68$ are at least four times larger than their LoS counterparts. Please see Tables \ref{tab:PL} and \ref{tab:sha} for detailed results.

We note that the confidence intervals (CIs) for the NLoS case span a much larger range than for the LoS case. This is to a small degree caused by a smaller collection of NLoS points when compared to LoS points (17 vs 21), but is mostly due to the larger spread of the actually measured results. We also note that the number of measurements and spreads is - from visual inspection - in line with other measurement campaigns at high frequencies in the literature, but since they generally do not provide CIs for their parameter fits, a comparison of CIs is not possible.

\begin{table*}[t!]
\centering
\caption{Path loss parameters with $95\%$ confidence interval.}
\label{tab:PL}
{%
\begin{tabular}{|c|c|c|c|c|c|c|c|}
\hline
\multicolumn{1}{|c|}{\multirow{2}{*}{\textbf{Parameter}}} & \multicolumn{6}{c|}{\textbf{Linear model parameters estimated with 95\% CI}} \\ \cline{2-7} 
\multicolumn{1}{|l|}{} & \multicolumn{1}{c|}{$\alpha$} & \multicolumn{1}{c|}{$\alpha_{min,95\%}$} & \multicolumn{1}{c|}{$\alpha_{max,95\%}$} & \multicolumn{1}{c|}{$\beta$} & \multicolumn{1}{c|}{$\beta_{min,95\%}$} & \multicolumn{1}{c|}{$\beta_{max,95\%}$} \\ \hline \hline
$PL_{Omni}^{LOS}$ & 76.77 & 75.5 & 78.05 & 1.74 & 1.65 & 1.83 \\ \hline
$PL_{Max-Dir}^{LOS}$ & 76.77 & 75.24 & 78.3 & 1.78 & 1.67 & 1.89 \\ \hline
$PL_{Omni}^{LOS} OLS$ &  76.53 & 74.74 & 78.32 & 1.8 & 1.68 & 1.93 \\ \hline
$PL_{Max-Dir}^{LOS} OLS$ & 76.42 & 74.27 & 78.86 & 1.86 & 1.71 & 2.01 \\ \hline
$PL_{Omni}^{NLOS}$ & 95.45 & 87.29 & 103.61 & 1.49 & 0.96 & 2.03 \\ \hline
$PL_{Max-Dir}^{NLOS}$ & 100.47 & 89.78 & 111.16 & 1.35 & 0.65 & 2.06 \\ \hline
$PL_{Omni}^{NLOS} OLS$ & 96.26 & 85.12 & 107.41 & 1.53 & 0.81 & 2.25 \\ \hline
$PL_{Max-Dir}^{NLOS} OLS$ & 101.03 & 85.97 & 116.08 & 1.45 & 0.48 & 2.42 \\ \hline
\end{tabular}%
}
\end{table*}

\begin{table*}[t!]
\centering
\caption{Shadowing model parameters with $95\%$ confidence interval.}
\label{tab:sha}
{%
\begin{tabular}{|c|c|c|c|c|c|c|}
\hline
\multicolumn{1}{|c|}{\multirow{2}{*}{\textbf{Parameter}}} & \multicolumn{6}{c|}{\textbf{Statistical model parameters estimated with 95\% CI}} \\ \cline{2-7} 
\multicolumn{1}{|l|}{} & \multicolumn{1}{c|}{$\mu$} & \multicolumn{1}{c|}{$\mu_{min,95\%}$} & \multicolumn{1}{c|}{$\mu_{max,95\%}$} & \multicolumn{1}{c|}{$\sigma$} & \multicolumn{1}{c|}{$\sigma_{min,95\%}$} & \multicolumn{1}{c|}{$\sigma_{max,95\%}$} \\ \hline \hline
$\epsilon_{Omni}^{LOS}$ & 0.58 & -0.08 & 1.24 & 1.45 & 1.11 & 2.09 \\ \hline
$\epsilon_{Max-Dir}^{LOS}$ & 0.8 & -0.02 & 1.63 & 1.81 & 1.38 & 2.61 \\ \hline
$\epsilon_{Omni}^{LOS} OLS$ & -0.01 & -0.65 & 0.62 & 1.4 & 1.07 & 2.02 \\ \hline
$\epsilon_{Max-Dir}^{LOS} OLS$ & 0.05 & -0.74 & 0.84 & 1.73 & 1.33 & 2.5 \\ \hline
$\epsilon_{Omni}^{NLOS}$ & 1.37 & -1.31 & 4.05 & 5.21 & 3.88 & 7.93 \\ \hline
$\epsilon_{Max-Dir}^{NLOS}$ & 2.09 & -1.37 & 5.54 & 6.72 & 5 & 10.23 \\ \hline
$\epsilon_{Omni}^{NLOS} OLS$ & 0.02 & -2.65 & 2.7 & 5.2 & 3.88 & 7.92 \\ \hline
$\epsilon_{Max-Dir}^{NLOS} OLS$ & 0 & -3.43 & 3.44 & 6.69 & 4.98 & 10.18 \\ \hline
\end{tabular}%
}
\end{table*}

\subsection{RMS delay spread}

The RMS delay spread in the LoS case ranges between 1 and 10 ns for the max-dir case, and 10-100 ns for the omni-directional case. More precisely, the average difference between the omni-directional and directional cases is approximately 7 dBs (i.e. 5 times on a linear scale). This significant difference follows from the fact that the narrow beamwidth of the horn antennas suppresses many significant reflected components, thus making the LoS component more dominant, and reducing the second central moment of the PDP. It is noteworthy that even for the most extreme case of max-dir delay spread in LoS situations, the RMS delay spread is on the order of several nanoseconds. Considering the extremely high data rates envisioned for THz, equalizers or equivalent structures will be required in many cases.  

\begin{figure*}[t!]
    \centering
    \begin{subfigure}[b]{0.45\textwidth}
        \centering
        \includegraphics[width=1\columnwidth]{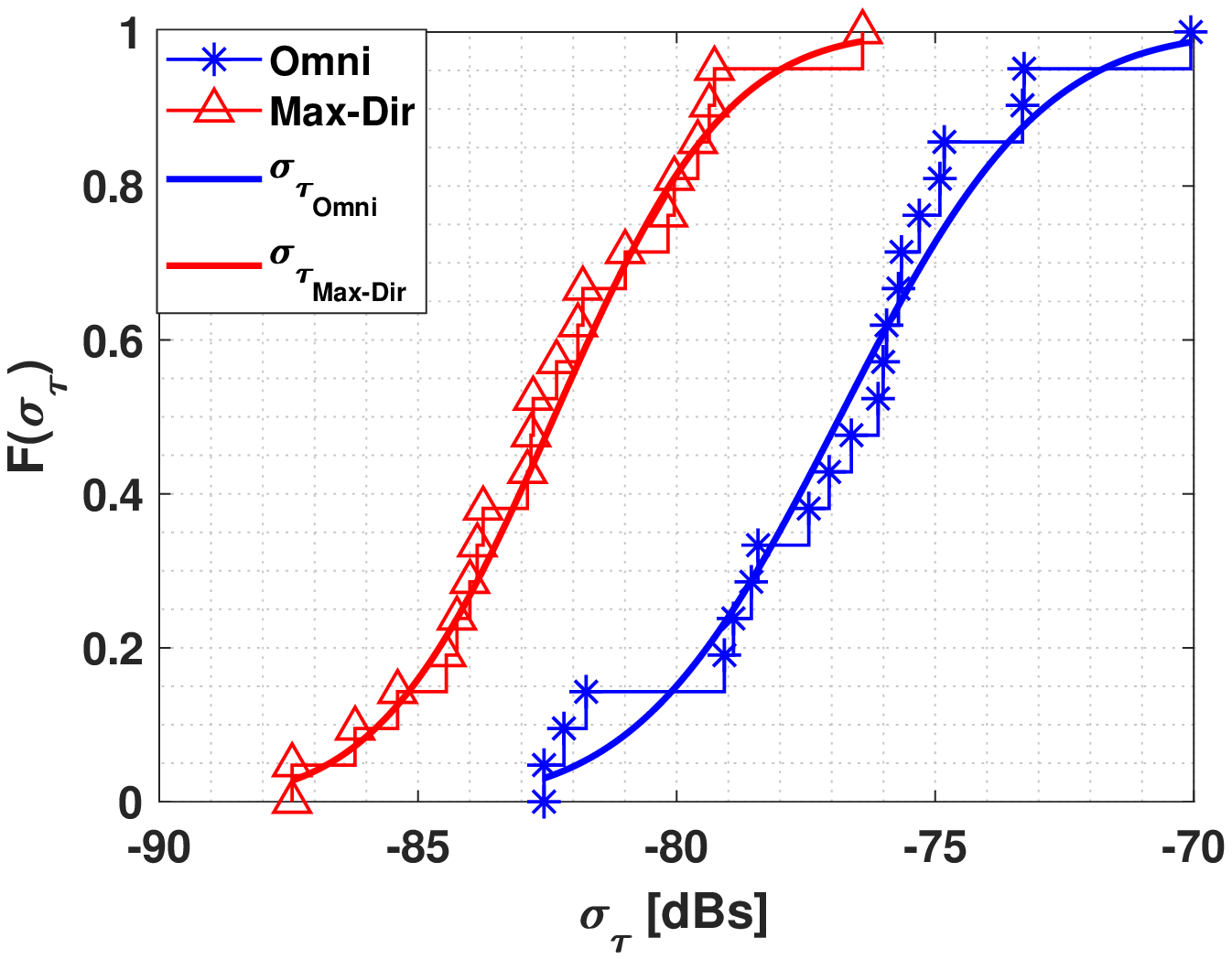}
        \caption{CDF of delay spread.}
        \vspace*{0mm}
        \label{fig:RMSDS-LOS-CDF}%
    \end{subfigure}
\begin{subfigure}[b]{0.45\textwidth}
        \centering
        \includegraphics[width=1\columnwidth]{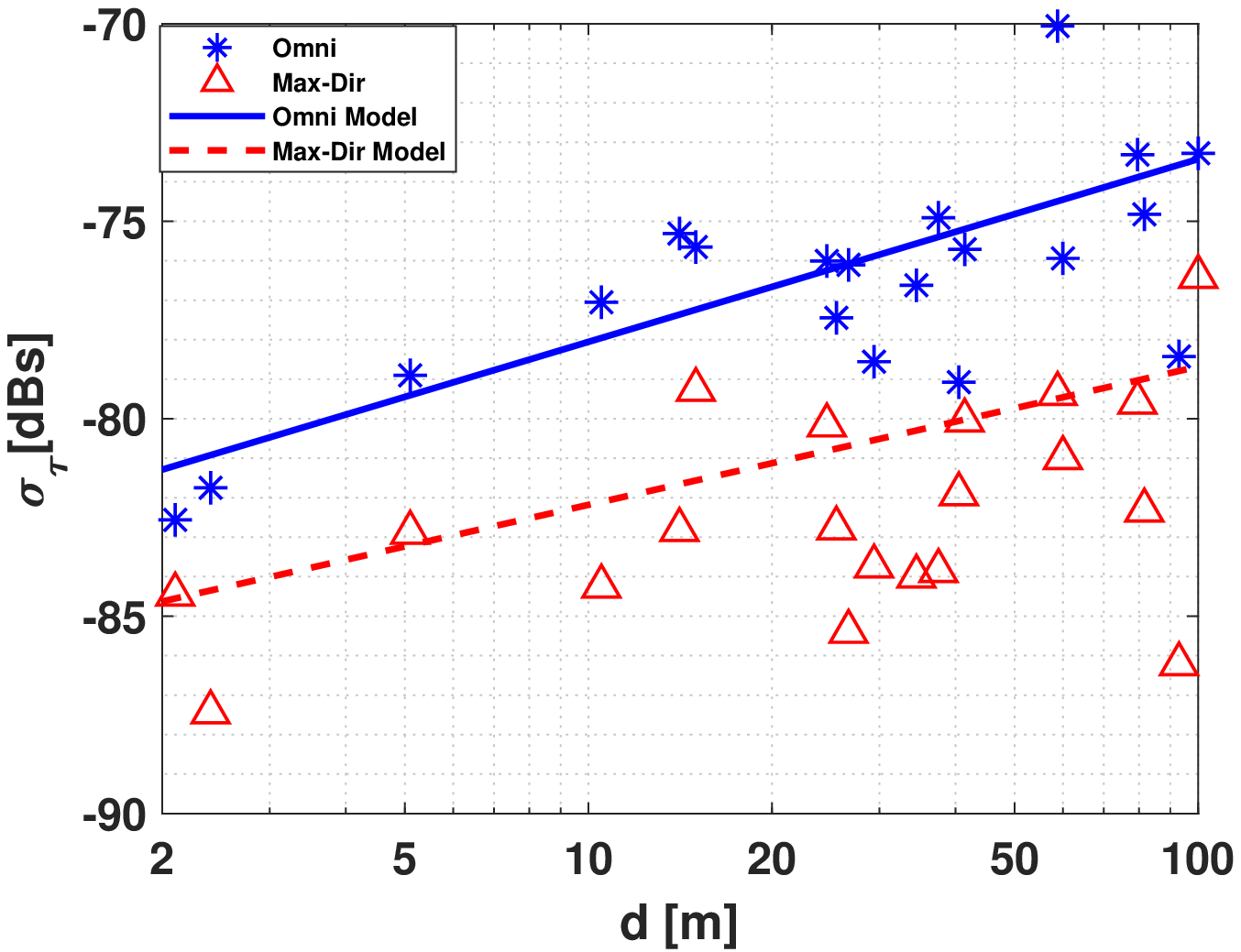}
        \caption{Linear modeling of $\sigma_\tau$ with weighting.}
        \vspace*{0mm}
        \label{fig:RMSDS-LOS-LF}%
    \end{subfigure}
\caption{Modeling of delay spread for LoS cases.}%
\vspace{-0 mm}
\label{fig:RMSDS-LOS}%
\vspace{-5mm}
\end{figure*}

When analyzing the cumulative distribution function (CDF), we find a similar behavior for both omni and max-dir cases: a lognormal distribution (i.e. Gaussian on a dB scale) as first suggested by Greenstein \cite{580786} and observed for mmWave in \cite{umit}, provides an excellent fit, as shown in Fig. \ref{fig:RMSDS-LOS} (a). The standard deviation for the omni-directional and directional case are similar ($\sigma_{\tau_{Omni}}=3.05$,$\sigma_{\tau_{Max-Dir}}=3.08$). Note that giving delay spreads in units of dBs is common in the channel modeling literature and is used, e.g., in the 3GPP channel model, since it is a natural fit to the lognormal model for the delay spread cdf. 

Delay spread shows a linear relationship with respect to the distance based on our analysis of the various measurements.  
In the LoS cases, the delay spread increases with the distance between the Tx and Rx for both the directional and omni-directional case; the CI in this case encompasses only positive slopes. The increase with distance occurs
because at large distances, a larger number MPCs with power comparable to the LoS MPC appear. This is also consistent with observations of various measurement campaigns at lower frequencies. All of these observations can be seen in Fig. \ref{fig:RMSDS-LOS} (b).

\begin{figure*}[t!]
    \centering
\begin{subfigure}[b]{0.45\textwidth}
        \centering
        \includegraphics[width=1\columnwidth]{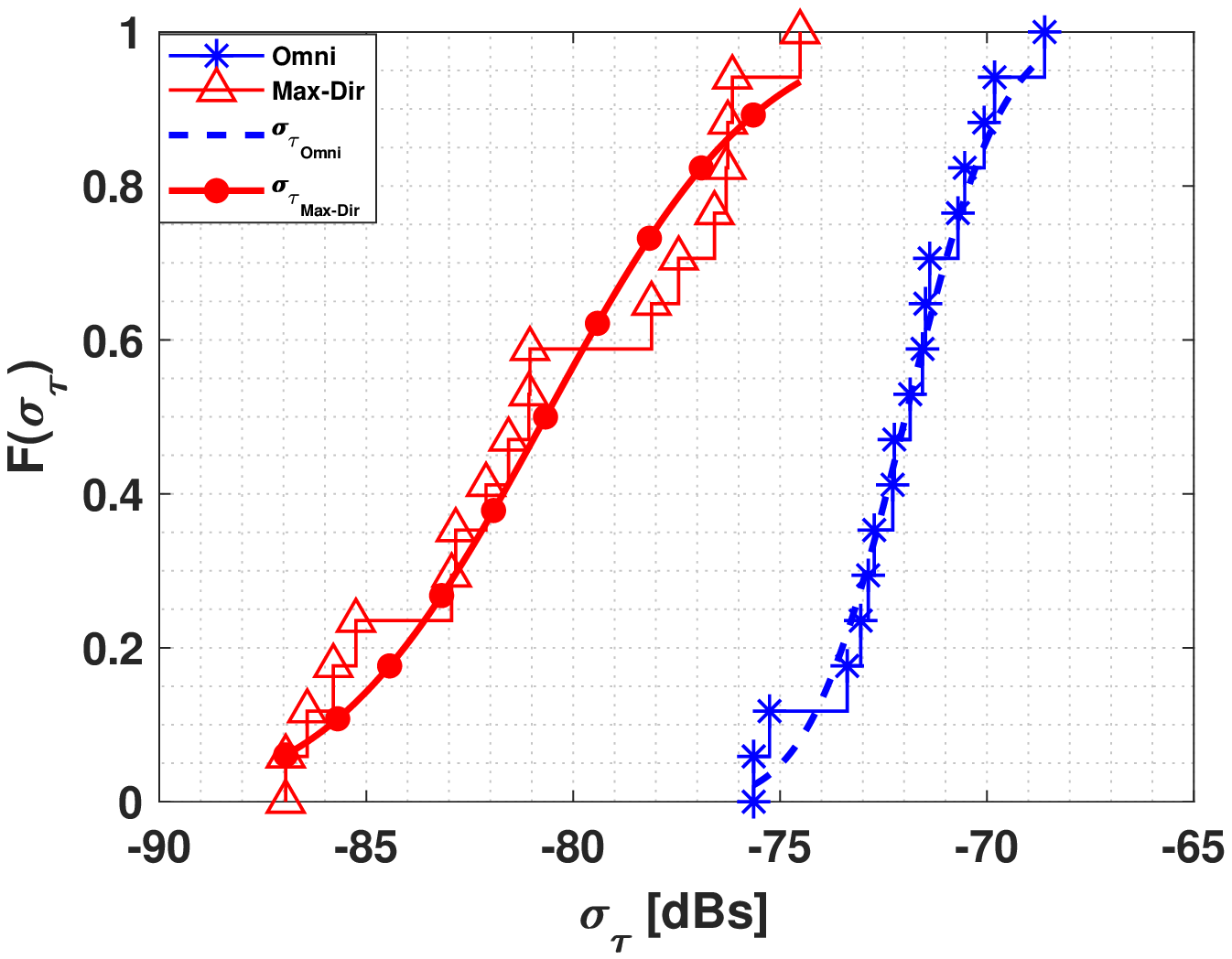}
        \caption{CDF}
        \vspace*{0mm}
        \label{fig:RMSDS-NLOS-CDF}%
    \end{subfigure}
     \begin{subfigure}[b]{0.45\textwidth}
        \centering
        \includegraphics[width=1\columnwidth]{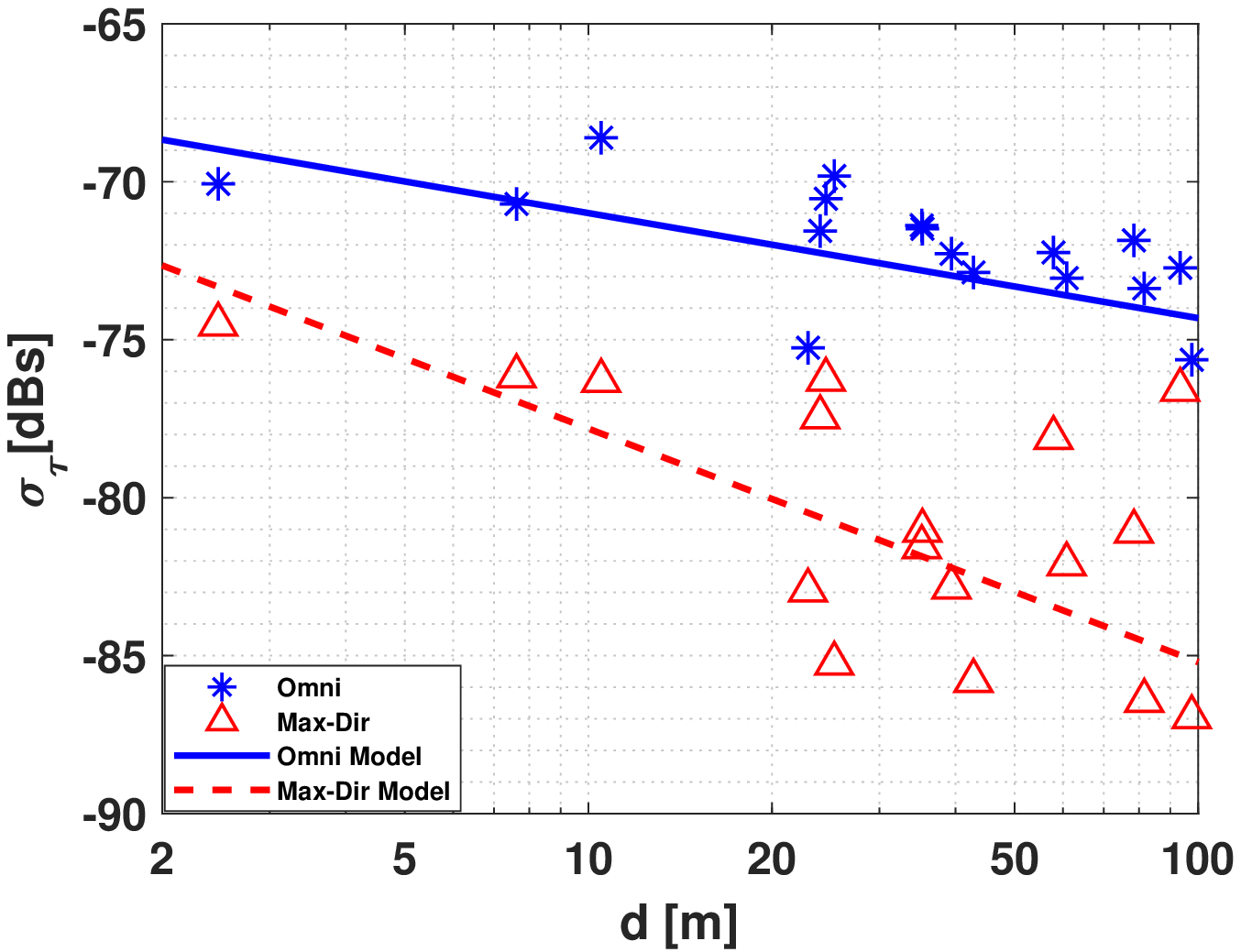}
        \caption{Linear fitting with $log_{10}(d)$ weighting.}
        \vspace*{0mm}
        \label{fig:RMSDS-NLOS-LF}%
    \end{subfigure}
\caption{Modeling of delay spread for NLoS points.}%
\vspace{-0 mm}
\label{fig:RMSDS-NLOS}%
\vspace{-5mm}
\end{figure*}

For the NLoS case, we expect larger delay spread compared to the LoS since the absence of a single dominant component increases the second central moment. This is indeed borne out by the measurements in Fig. \ref{fig:RMSDS-NLOS} (a) where delay spread values of up to 30ns occur in the max-dir case. For the omni case, while the maximum observed delay spread remains at 150 ns, the average value of around 65 ns is about a factor of 3 higher than in the LoS case (20 ns). The variance observed in the directional case is larger ($\sigma_{\tau_{Omni}}=1.8$,$\sigma_{\tau_{Max-Dir}}=3.96$). 

Furthermore, - in contrast to the LoS case - in the directional case the delay spread decreases with increasing distance; the CI only contains negative values, and the the absolute value of the slope is larger than in the LoS case. Given the small number of significant MPCs arriving at the Rx, we note that the variance might change considerably depending on the measurement scenario. For the omni-directional case, the delay spread also decreases with increasing distance, but in a less pronounced way, and the CI contains both positive and negative values. This can be explained by the fact that more MPCs are added at longer distances. The linear fittings are shown in Fig. \ref{fig:RMSDS-NLOS} (b). 

We summarize the detailed results for various parameters in Tables \ref{tab:linear-model-RMSDS} and \ref{tab:RMSDS_CDF}.

\begin{table*}[t!]
\centering
\caption{Linear model parameters for $\sigma_{\tau}$ with $95\%$ confidence interval.}
\label{tab:linear-model-RMSDS}
{%
\begin{tabular}{|c|c|c|c|c|c|c|c|}
\hline
\multicolumn{1}{|c|}{\multirow{2}{*}{\textbf{Parameter}}} & \multicolumn{6}{c|}{\textbf{Linear model parameters estimated with 95\% CI}} \\ \cline{2-7} 
\multicolumn{1}{|l|}{} & \multicolumn{1}{c|}{$\alpha$} & \multicolumn{1}{c|}{$\alpha_{min,95\%}$} & \multicolumn{1}{c|}{$\alpha_{max,95\%}$} & \multicolumn{1}{c|}{$\beta$} & \multicolumn{1}{c|}{$\beta_{min,95\%}$} & \multicolumn{1}{c|}{$\beta_{max,95\%}$} \\ \hline \hline
$\sigma_{\tau_{Omni}}^{LOS}$ & -82.68 & -84.1 & -81.26 & 4.63 & 3.62 & 5.64 \\ \hline
$\sigma_{\tau_{Max-Dir}}^{LOS}$ & -85.68 & -88.2 & -83.16 & 3.5 & 1.7 & 5.3 \\ \hline
$\sigma_{\tau_{Omni}}^{NLOS}$ & -67.67 & -68 & -65.34 & -3.32 & -4.86 & -1.79 \\ \hline
$\sigma_{\tau_{Max-Dir}}^{NLOS}$ & -70.43 & -74.7 & -66.15 & -7.39 & -10.21 & -4.57 \\ \hline
\end{tabular}%
}
\end{table*}

\begin{table*}[t!]
\centering
\caption{Statistical model parameters for $\sigma_{\tau}$ with $95\%$ confidence interval.}
\label{tab:RMSDS_CDF}
{%
\begin{tabular}{|c|c|c|c|c|c|c|}
\hline
\multicolumn{1}{|c|}{\multirow{2}{*}{\textbf{Parameter}}} & \multicolumn{6}{c|}{\textbf{Statistical model parameters estimated with 95\% CI}} \\ \cline{2-7} 
\multicolumn{1}{|l|}{} & \multicolumn{1}{c|}{$\mu$} & \multicolumn{1}{c|}{$\mu_{min,95\%}$} & \multicolumn{1}{c|}{$\mu_{max,95\%}$} & \multicolumn{1}{c|}{$\sigma$} & \multicolumn{1}{c|}{$\sigma_{min,95\%}$} & \multicolumn{1}{c|}{$\sigma_{max,95\%}$} \\ \hline \hline
$\sigma_{\tau_{Omni}}^{LOS}$ & -76.84 & -78.23 & -75.46 & 3.05 & 2.33 & 4.41 \\ \hline
$\sigma_{\tau_{Max-Dir}}^{LOS}$ & -82.37 & -83.57 & -81.16 & 2.64 & 2.02 & 3.81 \\ \hline
$\sigma_{\tau_{Omni}}^{NLOS}$ & -71.97 & -72.9 & -71.03 & 1.82 & 1.35 & 2.77 \\ \hline
$\sigma_{\tau_{Max-Dir}}^{NLOS}$ & -80.67 & -82.75 & -78.58 & 4.06 & 3.02 & 6.17 \\ \hline
\end{tabular}%
}
\end{table*}

\subsection{Angular spread}
For the statistics of the angular spread, a distinction between Tx and Rx is moot, since both Tx and Rx are at the same height (our focus is on D2D scenarios), and the selection of a node as Tx and Rx is arbitrary. Therefore, we can analyze the ensemble of all transceiver locations, without distinction of Tx/Rx (note that this would not be correct in a scenario where one node shows distinctive features, such as an access point in an elevated location). 

\begin{figure*}[t!]
    \centering
    \begin{subfigure}[b]{0.45\textwidth}
       \centering
        \includegraphics[width=\columnwidth]{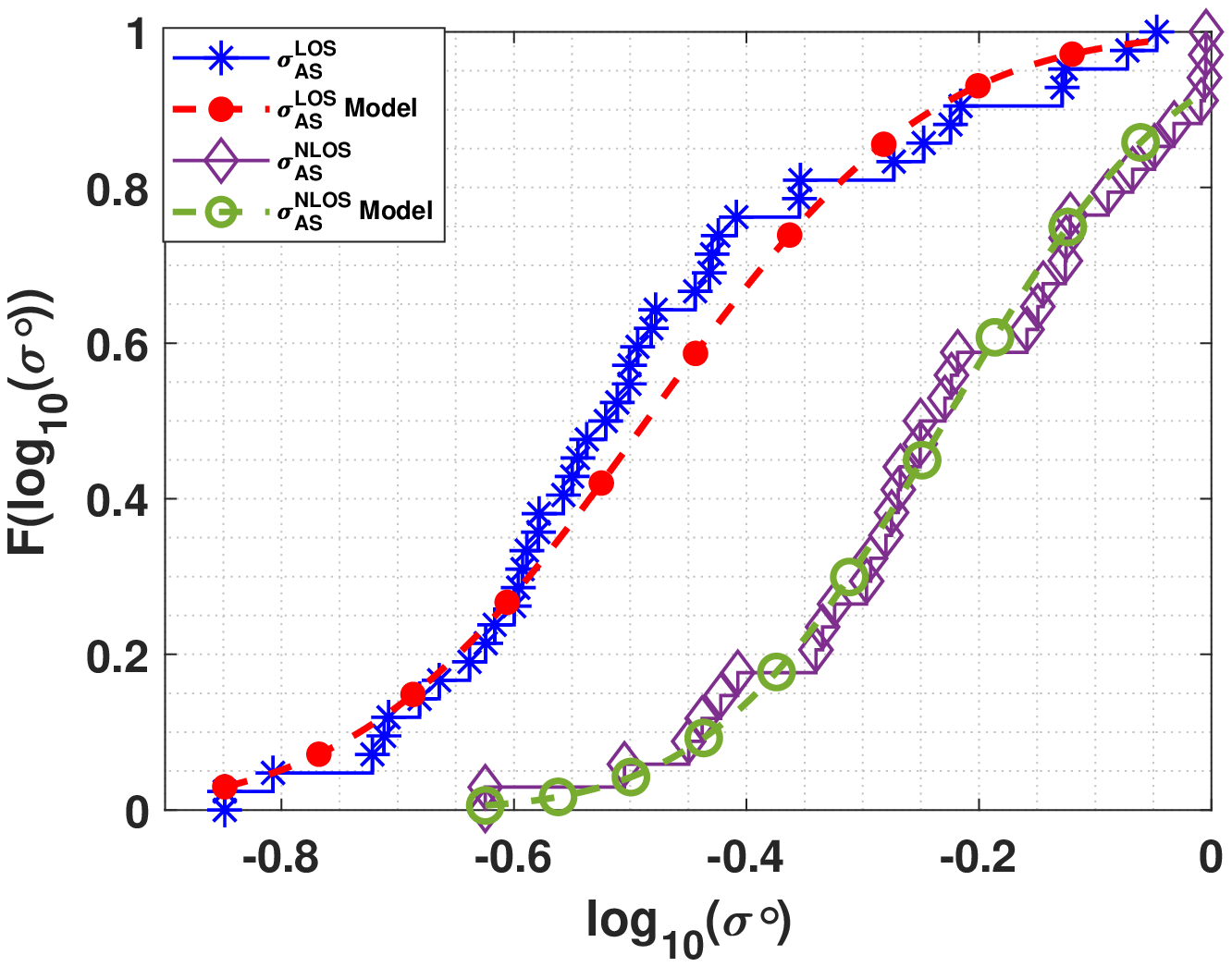}
        \caption{CDF}
        \vspace*{0mm}
        \label{fig:AS_CDF}%
       
    \end{subfigure}
\begin{subfigure}[b]{0.45\textwidth}
               \centering
        \includegraphics[width=\columnwidth]{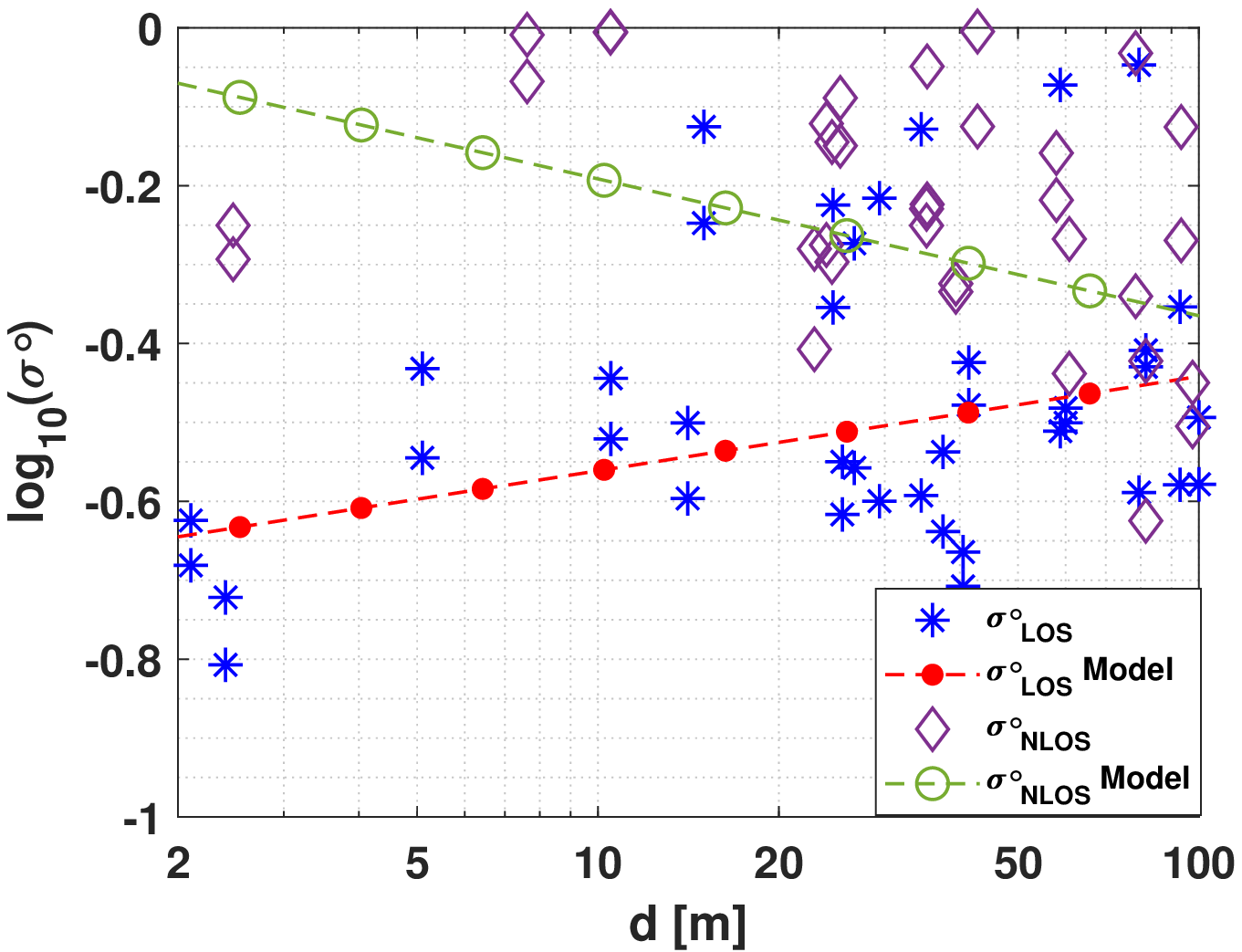}
        \caption{Linear fitting with $log_{10}(d)$ weighting.}
        \vspace*{0mm}
        \label{fig:AS_LINEAR}%
    \end{subfigure}
\caption{Modeling of $\sigma^\circ$ for all points.}%
\vspace{-0 mm}
\label{fig:AS}%
\vspace{-5mm}
\end{figure*}

As expected, LoS points show considerably lower values of angular spread compared to the NLoS points because of the presence of strong LoS components. However, as discussed in Sec. IV.B, there are also NLoS scenarios with low angular spread. This occurs in particular in street canyon scenarios that have concentrated beams, and therefore, a reduced angular span that leads to a decrease in the angular spread, since most of the components come from reflections inside the canyon. Fig. \ref{fig:AS} (a) demonstrates these observations. Please note that LoS values are significantly more likely - but not guaranteed - to have lower angular spread compared to the NLoS points. However, as the distance increases between the Tx and Rx, the angular spread increases as well because the difference of power between the LoS component and the rest of the MPCs is reduced. The opposite effect is observed for the NLoS points, though with larger spread values.

Linear fitting of angular spread versus distance is shown in Fig. \ref{fig:AS} (b). The angular spread increases for LoS points and decreases for NLoS situations, with increasing distance. These observations and their relevant causes are similar to those from Fig. \ref{fig:AS} (a). Detailed results for this case are presented in Tables \ref{tab:linear-model-AS} and \ref{tab:stat-model-AS}.

\begin{table*}[t!]
\centering
\caption{Linear model parameters for $\sigma^\circ$ with $95\%$ confidence interval.}
\label{tab:linear-model-AS}
{%
\begin{tabular}{|c|c|c|c|c|c|c|c|}
\hline
\multicolumn{1}{|c|}{\multirow{2}{*}{\textbf{Parameter}}} & \multicolumn{6}{c|}{\textbf{Linear model parameters estimated with 95\% CI}} \\ \cline{2-7} 
\multicolumn{1}{|l|}{} & \multicolumn{1}{c|}{$\alpha$} & \multicolumn{1}{c|}{$\alpha_{min,95\%}$} & \multicolumn{1}{c|}{$\alpha_{max,95\%}$} & \multicolumn{1}{c|}{$\beta$} & \multicolumn{1}{c|}{$\beta_{min,95\%}$} & \multicolumn{1}{c|}{$\beta_{max,95\%}$} \\ \hline \hline
$\sigma^\circ_{LOS}$ & -0.68 & -0.78 & -0.58 & 0.12 & 0.05 & 0.19 \\ \hline
$\sigma^\circ_{NLOS}$ & -0.02 & -0.17 & 0.13 & -0.17 & -0.27 & -0.08\\ \hline
\end{tabular}%
}
\end{table*}
    
\begin{table*}[t!]
\centering
\caption{Statistical model parameters for $\sigma^\circ$ with $95\%$ confidence interval.}
\label{tab:stat-model-AS}
{%
\begin{tabular}{|c|c|c|c|c|c|c|}
\hline
\multicolumn{1}{|c|}{\multirow{2}{*}{\textbf{Parameter}}} & \multicolumn{6}{c|}{\textbf{Statistical model parameters estimated with 95\% CI}} \\ \cline{2-7} 
\multicolumn{1}{|l|}{} & \multicolumn{1}{c|}{$\mu$} & \multicolumn{1}{c|}{$\mu_{min,95\%}$} & \multicolumn{1}{c|}{$\mu_{max,95\%}$} & \multicolumn{1}{c|}{$\sigma$} & \multicolumn{1}{c|}{$\sigma_{min,95\%}$} & \multicolumn{1}{c|}{$\sigma_{max,95\%}$} \\ \hline \hline
$\sigma^{\circ}_{LOS}$ & -0.49 & -0.55 & -0.43 & 0.19 & 0.16 & 0.25 \\ \hline
$\sigma^{\circ}_{NLOS}$ & -0.23 & -0.29 & -0.17 & 0.16 & 0.13 & 0.21 \\ \hline
\end{tabular}%
}
\end{table*}
    
\subsection{Power distribution of MPCs}
\begin{figure*}[t!]
    \centering
    \begin{subfigure}[b]{0.45\textwidth}
       \centering
        \includegraphics[width=1\columnwidth]{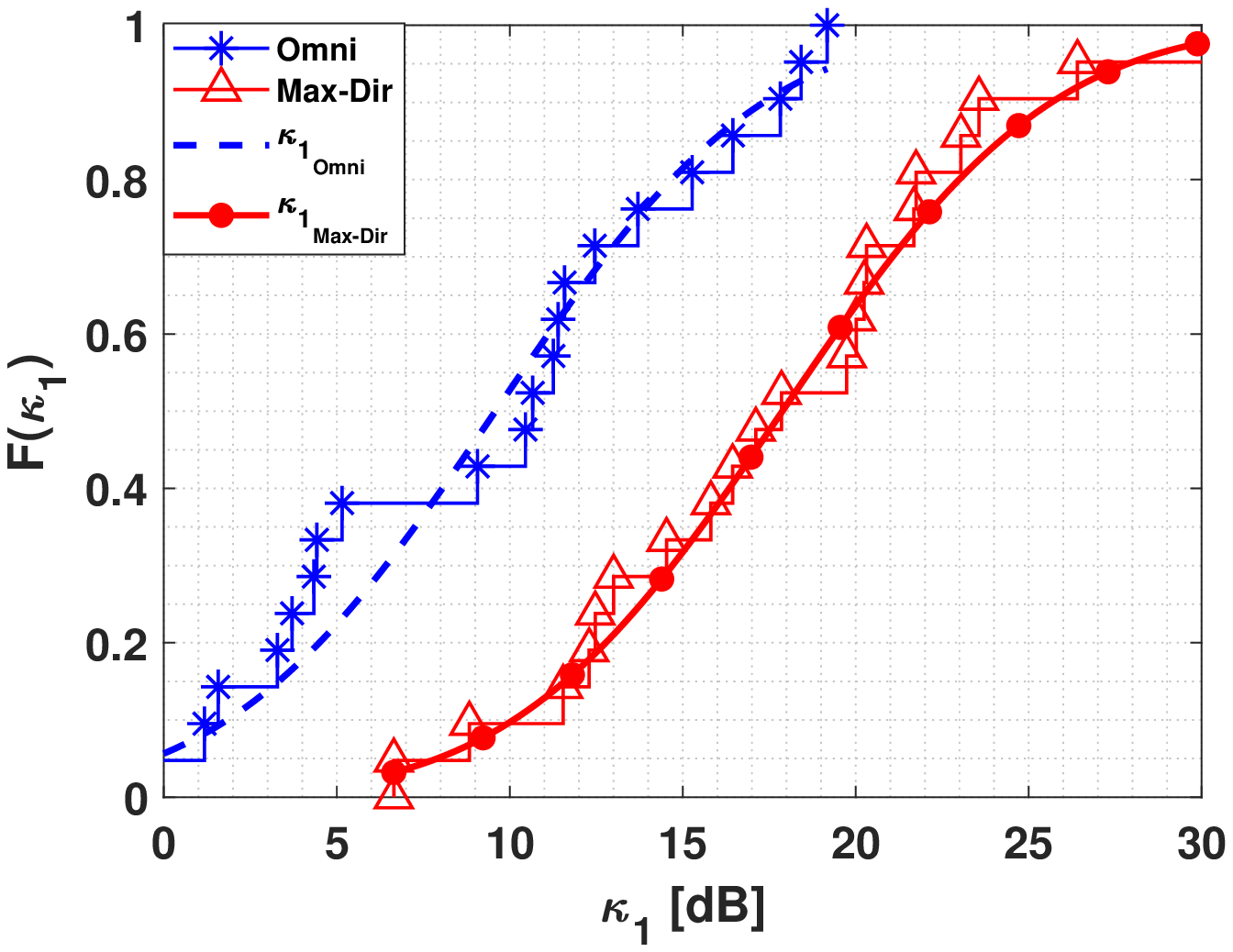}
        \caption{CDF.}
        \vspace*{0mm}
        \label{fig:k1_CDF}%

    \end{subfigure}
\begin{subfigure}[b]{0.45\textwidth}
        \centering
        \includegraphics[width=1\columnwidth]{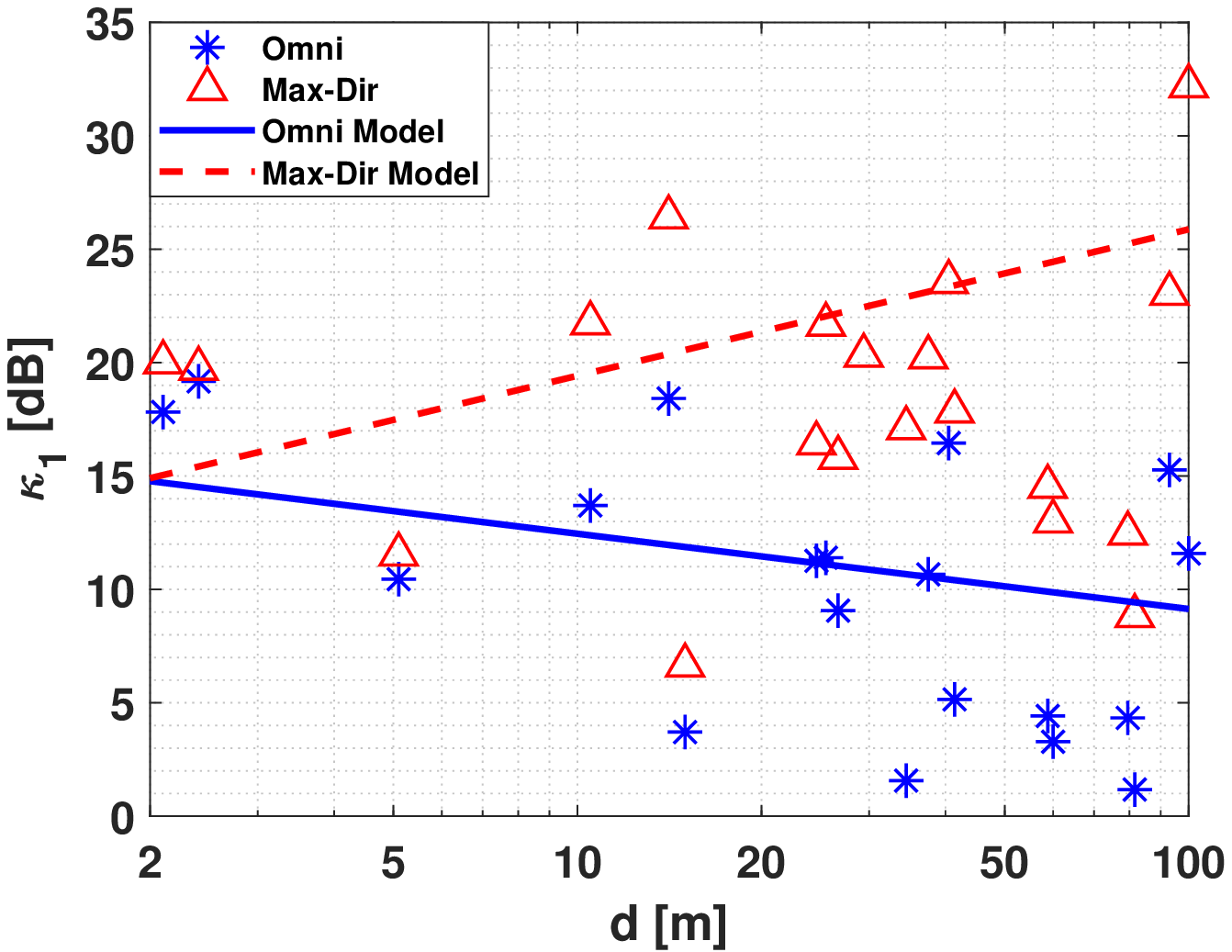}
        \caption{Linear fitting with $log_{10}(d)$ weighting.}
        \vspace*{0mm}
        \label{fig:k1_LS}%
    \end{subfigure}
\caption{Modeling of $\kappa_1$ for LoS points.}%
\vspace{-0 mm}
\label{fig:k1}%
\vspace{-5mm}
\end{figure*}

From physical considerations of the urban scenarios, we expect the values of $\kappa_1$ for directional cases to be larger than the omni-directional ones because of the large number of MPCs in the PDPs for the latter case. This is indeed borne out by the measurements: for the LoS case we find a directional mean of 18 dB, while the omni-directional mean is around 10 dB. The distributions of the $\kappa_1$ values follows a lognormal distribution, with a standard deviation is 6 dB in both cases as shown in Fig. \ref{fig:k1_CDF}. 

The omni-directional $\kappa_1$  shows a decrease as the distance increases because the reflected MPCs have powers closer to the strongest component. On the other hand, the directional cases shows an increasing trend as the distance between Tx and Rx increases because the spatial filtering provided by the horn antennas - the directional range to/from which the MPCs come is increased, but the horns filters out MPCs outside an angular range that is the beamwidth around the LoS direction. This is demonstrated in Fig. \ref{fig:k1} (b), where we observe the linear fitting for the parameter in LoS points.

\begin{figure*}[t!]
    \centering
\begin{subfigure}[b]{0.45\textwidth}
        \centering
        \includegraphics[width=1\columnwidth]{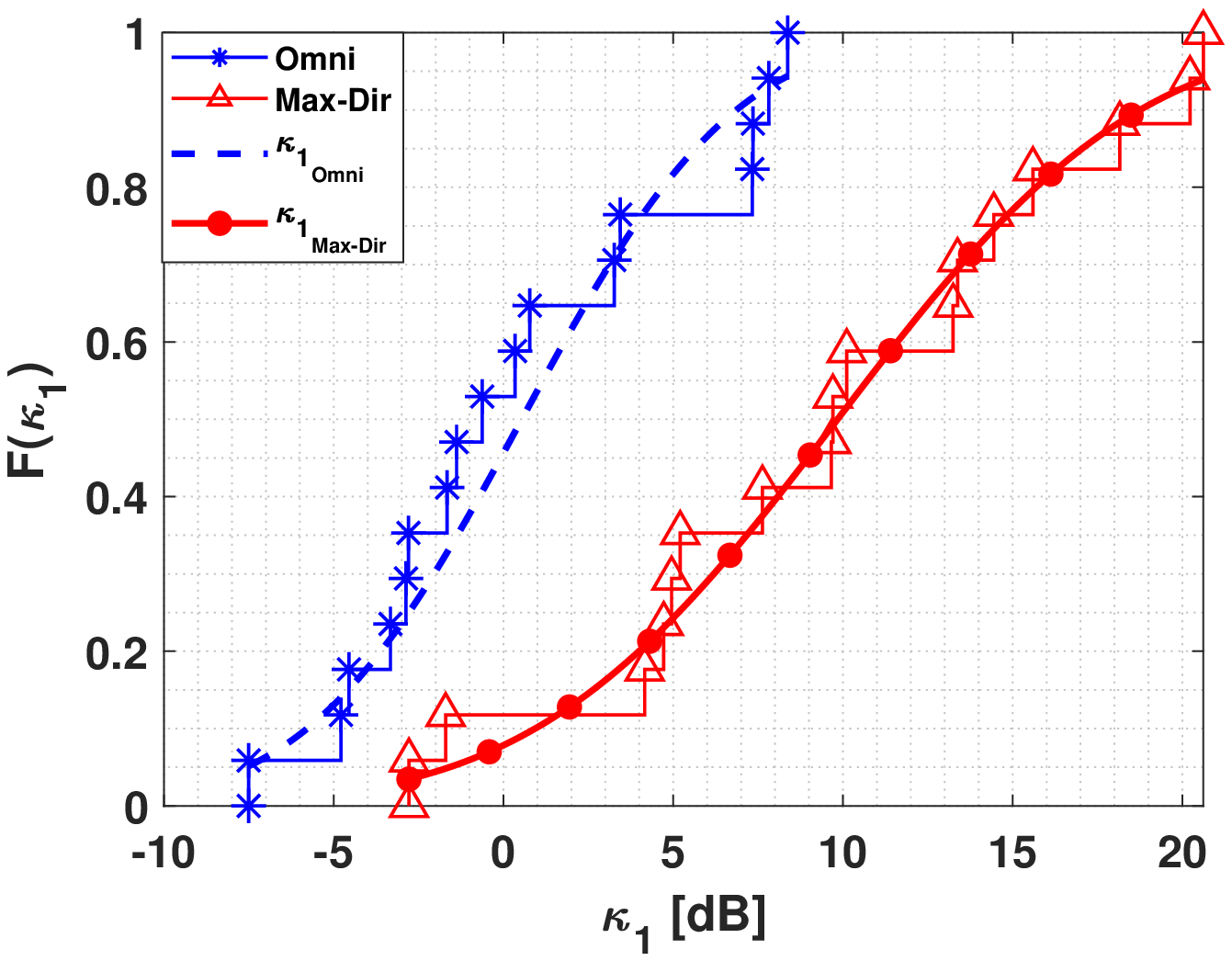}
        \caption{CDF.}
        \vspace*{0mm}
        \label{fig:k1_CDF_NLOS}%
    \end{subfigure}
        \begin{subfigure}[b]{0.45\textwidth}
        \centering
        \includegraphics[width=1\columnwidth]{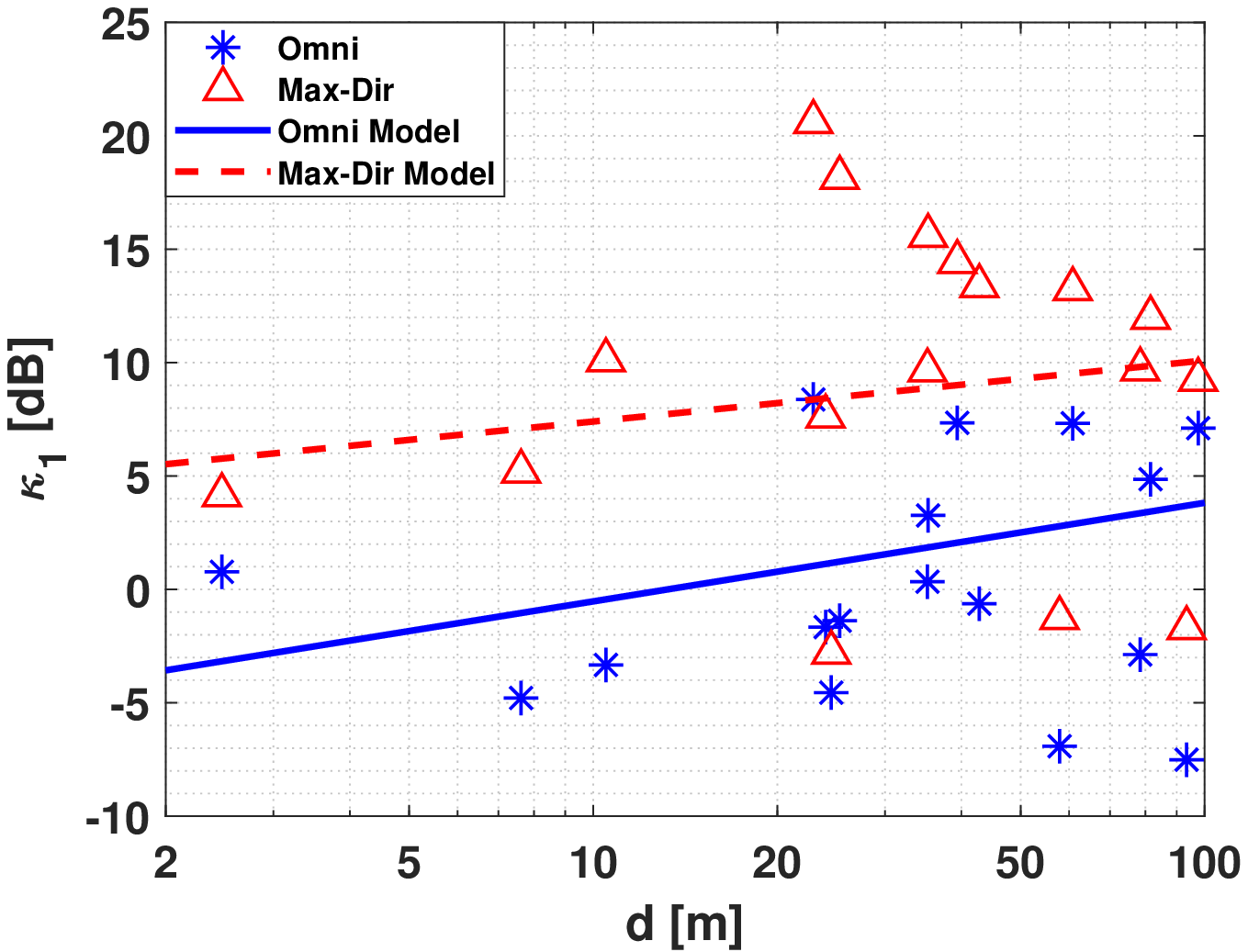}
        \caption{Linear fitting with $log_{10}(d)$ weighting.}
        \vspace*{0mm}
        \label{fig:k1_LS_NLOS}%
    \end{subfigure}
\caption{Modeling of $\kappa_1$ for NLoS points.}%
\vspace{-0 mm}
\label{fig:k1_NLOS}%
\vspace{-5mm}
\end{figure*}

Fig. \ref{fig:k1_NLOS} (a) and (b) show our observations for the NLoS measurements. The average $\kappa_1$ values are considerably lower, around 9 dB for the max-dir case, and 8 dB for the omni case; the standard deviation is 5 dB and 7 dB, respectively.

The omni-directional case also shows an increase with respect to distance as shown in Fig. \ref{fig:k1_NLOS} (b). This is an effects of the street canyon environment on some locations (e.g. Annenberg). More details of the fitting results can found in Tables \ref{tab:linear-model-kappa} and \ref{tab:stat-model-kappa}. 

\begin{table*}[t!]
\centering
\caption{Linear model parameters for $\kappa_1$ with $95\%$ confidence interval.}
\label{tab:linear-model-kappa}
{%
\begin{tabular}{|c|c|c|c|c|c|c|c|}
\hline
\multicolumn{1}{|c|}{\multirow{2}{*}{\textbf{Parameter}}} & \multicolumn{6}{c|}{\textbf{Linear model parameters estimated with 95\% CI}} \\ \cline{2-7} 
\multicolumn{1}{|l|}{} & \multicolumn{1}{c|}{$\alpha$} & \multicolumn{1}{c|}{$\alpha_{min,95\%}$} & \multicolumn{1}{c|}{$\alpha_{max,95\%}$} & \multicolumn{1}{c|}{$\beta$} & \multicolumn{1}{c|}{$\beta_{min,95\%}$} & \multicolumn{1}{c|}{$\beta_{max,95\%}$} \\ \hline \hline
$\kappa_{1_{Omni}}^{LOS}$ & 14.95 & 9.94 & 19.96 & -3.39 & -7.05 & 0.25\\ \hline
$\kappa_{1_{Max-Dir}}^{LOS}$ & 11.47 & 4.37 & 18.57 & 6.19 & 1 & 11.38 \\ \hline
$\kappa_{1_{Omni}}^{NLOS}$ & -4.88 & -11.97 & -2.22 & 4.35 & -0.33 & 9.03\\ \hline
$\kappa_{1_{Max-Dir}}^{NLOS}$ & 4.71 & -3.79 & 13.21 & 2.69 & -2.91 & 8.3 \\ \hline
\end{tabular}%
}
\end{table*}

\begin{table*}[t!]
\centering
\caption{Statistical model parameters for $\kappa_1$ with $95\%$ confidence interval.}
\label{tab:stat-model-kappa}
{%
\begin{tabular}{|c|c|c|c|c|c|c|}
\hline
\multicolumn{1}{|c|}{\multirow{2}{*}{\textbf{Parameter}}} & \multicolumn{6}{c|}{\textbf{Statistical model parameters estimated with 95\% CI}} \\ \cline{2-7} 
\multicolumn{1}{|c|}{} & \multicolumn{1}{c|}{$\mu$} & \multicolumn{1}{c|}{$\mu_{min,95\%}$} & \multicolumn{1}{c|}{$\mu_{max,95\%}$} & \multicolumn{1}{c|}{$\sigma$} & \multicolumn{1}{c|}{$\sigma_{min,95\%}$} & \multicolumn{1}{c|}{$\sigma_{max,95\%}$} \\ \hline \hline
$\kappa_{1_{Omni}}^{LOS}$ & 9.58 & 6.83 & 12.34 & 6.05 & 4.63 & 8.73 \\ \hline
$\kappa_{1_{Max-Dir}}^{LOS}$ & 17.88 & 15.12 & 20.64 & 6.07 & 4.64 & 8.76 \\ \hline
$\kappa_{1_{Omni}}^{NLOS}$ & 0.54 & -2 & 3.8 & 4.93 & 3.67 & 7.51 \\ \hline
$\kappa_{1_{Max-Dir}}^{NLOS}$ & 9.84 & 6.27 & 13.42 & 6.95 & 5.18 & 10.58 \\ \hline
\end{tabular}%
}
\end{table*}

\subsection{Summary of results}
\label{sec:summary}
In this section, we summarize the key results that may be very useful from a system design perspective. Table \ref{tab:linear-model-summary} presents the results of various linear fittings of the distance dependence of parameters for LoS and NLoS cases respectively. Table \ref{tab:stat-model-summary} provides the detailed results for various statistical fits done in the paper for the LoS and NLoS cases. These results can be directly used for system simulations.  

Finally, please note that while a larger number of channel measurements provides better statistical validity, the time required for doing measurement campaigns like the one presented here is significant (several months). For this reason, most initial channel models are based on a small number of measurement points to satisfy the need of the wireless community for a realistic basis for system design; of course the models can be subsequently improved as more measurements are conducted.

\begin{table*}[t!]
\centering
\caption{Linear model parameters summary.}
\label{tab:linear-model-summary}
{%
\begin{tabular}{|c|c|c|}
\hline
\multicolumn{1}{|c|}{\textbf{Parameter}} & \multicolumn{1}{c|}{$\alpha$} & \multicolumn{1}{c|}{$\beta$} \\ \hline \hline
$PL_{Omni}^{LOS}$ & 76.77 & 1.74 \\ \hline
$PL_{Max-Dir}^{LOS}$ & 76.77 & 1.78 \\ \hline
$PL_{Omni}^{LOS} OLS$ &  76.53 & 1.8 \\ \hline
$PL_{Max-Dir}^{LOS} OLS$ & 76.42 & 1.86 \\ \hline
$\sigma_{\tau_{Omni}}^{LOS}$ & -82.7 & 4.65 \\ \hline
$\sigma_{\tau_{Max-Dir}}^{LOS}$ & -86.96 & 4.26 \\ \hline
$\kappa_{1_{Omni}}^{LOS}$ & 14.95 & -3.39  \\ \hline
$\kappa_{1_{Max-Dir}}^{LOS}$ & 11.47 & 6.19 \\ \hline
$\sigma^\circ_{LOS}$ & -0.68 & 0.12\\ \hline
$\sigma^\circ_{NLOS}$ & -0.02 & -0.17\\ \hline
$PL_{Omni}^{NLOS}$ & 95.45 & 1.49 \\ \hline
$PL_{Max-Dir}^{NLOS}$ & 100.47 & 1.35 \\ \hline
$PL_{Omni}^{NLOS} OLS$ & 96.26 & 1.53 \\ \hline
$PL_{Max-Dir}^{NLOS} OLS$ & 101.03 & 1.45 \\ \hline
$\sigma_{\tau_{Omni}}^{NLOS}$ & -67.74 & -3.24\\ \hline
$\sigma_{\tau_{Max-Dir}}^{NLOS}$ & -70.77 & -6.96\\ \hline
$\kappa_{1_{Omni}}^{NLOS}$ & -4.88 & 4.35\\ \hline
$\kappa_{1_{Max-Dir}}^{NLOS}$ & 4.71 & 2.69 \\ \hline
\end{tabular}%
}
\end{table*}

\begin{table*}[t!]
\centering
\caption{Statistical model parameters summary.}
\label{tab:stat-model-summary}
{%
\begin{tabular}{|c|c|c|}
\hline
\multicolumn{1}{|c|}{\textbf{Parameter}} & \multicolumn{1}{c|}{$\mu$} & \multicolumn{1}{c|}{$\sigma$}\\ \hline \hline
$\epsilon_{Omni}^{LOS}$ & 0.58 & 1.45 \\ \hline
$\epsilon_{Max-Dir}^{LOS}$ & 0.8 & 1.81 \\ \hline
$\epsilon_{Omni}^{LOS} OLS$ & -0.01 & 1.4\\ \hline
$\epsilon_{Max-Dir}^{LOS} OLS$ & 0.05 & 1.73\\ \hline
$\sigma^{\circ}_{LOS}$ & -0.49 & 0.19\\ \hline
$\sigma_{\tau_{Omni}}^{LOS}$ & -76.84 & 3.05\\ \hline
$\sigma_{\tau_{Max-Dir}}^{LOS}$ & -83.15 & 3.08 \\ \hline
$\kappa_{1_{Omni}}^{LOS}$ & 9.58 & 6.05 \\ \hline
$\kappa_{1_{Max-Dir}}^{LOS}$ & 17.88 & 6.07 \\ \hline
$\epsilon_{Omni}^{NLOS}$ & 1.37 & 5.21 \\ \hline
$\epsilon_{Max-Dir}^{NLOS}$ & 2.09 & 6.72\\ \hline
$\epsilon_{Omni}^{NLOS} OLS$ & 0.02 & 5.2 \\ \hline
$\epsilon_{Max-Dir}^{NLOS} OLS$ & 0 & 6.69 \\ \hline
$\sigma^{\circ}_{NLOS}$ & -0.23 & 0.16 \\ \hline
$\sigma_{\tau_{Omni}}^{NLOS}$ & -71.92 & 1.8 \\ \hline
$\sigma_{\tau_{Max-Dir}}^{NLOS}$ & -80.34 & 3.96 \\ \hline
$\kappa_{1_{Omni}}^{NLOS}$ & 0.54 & 4.93 \\ \hline
$\kappa_{1_{Max-Dir}}^{NLOS}$ & 9.84 & 6.95\\ \hline
\end{tabular}%
}
\end{table*}

\section{Conclusions}



In this paper, we presented the results of the first extensive wideband, double-directional THz outdoor channel measurements for urban D2D scenarios. We provided an overview of the measurement methodology and environments, as well as the signal processing to extract parameters characterizing the channels. Most importantly, we provided a parameterized statistical description of our measurement results that can be used to assess THz systems. We can draw some first conclusions about the implications on system design and deployment:
\begin{itemize}
\item At 100 m distance, the THz channel exhibits a path loss of approximately 110 dB in LoS and 125 dB in NLoS. A simple link budget estimate shows that for a 1 GHz bandwidth, 23 dB gain antennas at both link ends, a required operating SNR of 5 dB, a receiver noise figure of 5 dB, and an assumed conducted transmit power of 10 dBm, communication remains possible even with a path loss of $10+46 +174-90-5-5=130$ dB. In other words, our results indicate that D2D communications over up to 100 m distance are feasible even in NLoS situations from a link budget point of view.
\item Contrary to the assumption in many theoretical papers, we observe rich multipath in a number of scenarios from the perspective of the current measurement system and the corresponding antennas. The spatial filtering that is inherent in the use of beam antennas (horns or phased arrays) helps to reduce the delay spread, but cannot eliminate it to the point where equalizers become redundant. For the max-dir case, the RMS delay spread (at the 90 \% level) is 5 ns and 20 ns. While the need for equalization might depend on the particular modulation scheme, these values indicate that a need for equalization cannot be dismissed a priori, even for the relatively low bandwidth (for THz) of 1 GHz. 
\item Another indication of rich multipath is the relatively large angular spread that is observed for both LoS and NLoS scenarios. This indicates that for multi-user systems, inter-user interference remains a significant problem even in the presence of large antenna arrays. Note that the effect of such spatial dispersion on signal-to-interference ratio might be compounded by the near-far effect: while MPCs that are 30 dB weaker than a LoS component might seem negligible, they can introduce significant inter-user interference in a nearby victim receiver that receives its own desired signal from a far-away transmitter. Thus, relying on the channel to provide a ``natural" separation of users in different directions might be optimistic. 
\item We note that for even for the max-dir case, the $\kappa_1$ that is guaranteed in 90 \% of all cases is only 0 dB for NLoS. Thus, even with directional antennas, it is not possible to rely on small fading depth in the design of THz transceivers.  
\end{itemize}

While these conclusions are mainly qualitative, they provide important insights into THz system behavior and show that some common conjectures that had been made about THz systems do not hold in the actually measured channels presented in this paper.  

Thus, the presented measurements should be of considerable interest not only for the propagation community, but more importantly for communication theorists and system designers. 
\section*{Acknowledgment}
The authors would like to thank the Semiconductor Research Corporation (SRC), whose ComSenTer project partially funded this work. Helpful discussions with Sundeep Rangan and Mark Rodwell are gratefully acknowledged. The work of AFM was also supported by the National Science Foundation. The work of JGP was supported by the Foreign Fulbright Ecuador SENESCYT Program.

\end{document}